\documentclass[preprint,preprintnumbers,amsmath,amssymb]{revtex4-1}

\usepackage{amsmath}
\usepackage{amssymb}
\usepackage{graphicx}% Include figure files
\usepackage{dcolumn}% Align table columns on decimal point
\usepackage{bm}% bold math
\usepackage{color}

\begin{document}
\title{Swarming and Pattern Formation due to Selective Attraction and Repulsion}% Force line breaks with \\

\author{Pawel Romanczuk} \email{prom@ppks.mpg.de} 
\affiliation{Max Planck Insitute for the Physics of Complex Systems, N{\"o}thnitzerstr. 38, 01187 Dresden, Germany}
\author{Lutz Schimansky-Geier}
\affiliation{Department of Physics, Humboldt Universit{\"a}t zu Berlin, Newtonstr 15, 12489 Berlin, Germany}
\affiliation{Instituto de F{\`i}sica Interdisciplinar y Sistemas Complejos (IFISC),
CSIC-UIB,Campus Universitat de les Illes Balears
E-07122 Palma de Mallorca, Spain}

\begin{abstract}
We discuss the collective dynamics of self-propelled particles with selective attraction and repulsion interactions. Each particle, or individual, may respond differently to its neighbors depending on the sign of their relative velocity. Thus, it is able to distinguish  approaching (coming closer) and moving away individuals. This differentiation of the social response is motivated by the response to looming visual stimuli and may be seen as a generalization of the previously proposed, biologically motivated, escape and pursuit interactions. The model can account for different types of behavior such as pure attraction, pure repulsion, or escape and pursuit depending on the values (signs) of the different response strengths, and provides, in the light of recent experimental results, an interesting alternative to previously proposed models of collective motion with an explicit velocity-alignment interaction.
We show the onset of large scale collective motion in a subregion of the parameter space, which corresponds to an effective escape and/or pursuit response. Furthermore, we discuss the observed spatial patterns and show how kinetic description of the dynamics can be derived from the individual based model.
   
\end{abstract}

\maketitle
\section{Introduction}

Collective motion in biology, as observed for example in flocks of birds, schools of fish or within bacterial colonies, is a fascinating display of natural self-organization. Over the years, it has been the topic of numerous scientific publications addressing it from very different angles and with different questions in mind, both experimentally and theoretically. From a more biological perspective the interesting questions are the evolutionary advantages and the biological and ecological function of collective behavior in various species \cite{couzin_collective_2002,krause_living_2002,yates_inherent_2009,bazazi_nutritional_2011}, whereas physicist addresses rather the question about universal laws and phase-transition behavior in minimal models of collective motion \cite{vicsek_novel_1995,toner_long-range_1995,toner_flocks_1998,gregoire_onset_2004,aldana_phase_2007,chate_collective_2008,romanczuk_collective_2009,romanczuk_active_2012}. The design, control and stability of collective dynamics in multi-agent systems is also a major research topic in engineering \cite{leonard_collective_2007,sepulchre_stabilization_2008,turgut_self-organized_2008,ferrante_flocking_2011}, and the general properties of related mathematical models are under active investigation in mathematics \cite{cucker_emergent_2007,carrillo_particle_2010}.  

Most of the mathematical models for collective motion proposed in the literature, contain some sort of explicit velocity-alignment mechanisms, which tends to align the velocity of a focal individual with the velocity of its neighbors \cite{niwa_self-organizing_1994,vicsek_novel_1995,couzin_collective_2002,couzin_effective_2005,romanczuk_beyond_2008,bode_making_2010}. However, recent experimental studies of collective behavior in fish do not find any clear evidence for the existence of an explicit velocity-alignment interaction \cite{herbert-read_inferring_2011,katz_inferring_2011}. 
Only relatively few models, have analyzed the onset of collective motion without an explicit alignment mechanisms based on purely repulsive and attractive interactions (see e.g. \cite{ebeling_swarms_2001,peruani_nonequilibrium_2006,grossman_emergence_2008,strefler_swarming_2008,strombom_collective_2011}). 
Recently, motivated by empirical evidence for cannibalism as the driving force of collective migration in certain insect species \cite{simpson_cannibal_2006,bazazi_collective_2008}, we have proposed a model of collective motion based on escape and pursuit responses \cite{romanczuk_collective_2009,bazazi_nutritional_2011}. In this escape-pursuit model, individuals are reacting to their neighbors by moving away from others approaching them from behind (escape), and/or increasing their velocity towards those who are moving away in front of them (pursuit). This kind of social response requires individuals to distinguish between approach and movement away as well as between individuals in front and behind them. This previous model can be considered to belong to a broader class of selective attraction-repulsion models, which we believe are very promising for theoretical modelling of collective motion in biology. Here we discuss and analyze a generalization of the original escape-pursuit model to the case where self-propelled agents (or particles) are responding selectively to approaching and moving individuals without taking their relative position into account \cite{guttal_cannibalism_2012}. Furthermore, we do not put any restrictions on the sign (direction) of the effective social forces modelling the selective response. This allows to account for different social behavior types, such as pure attraction, pure repulsion and escape and pursuit, in a single model, by the same set of social forces, only by changing the values of the response strengths.

We will start with the definition of the individual based model in terms of stochastic differential equations and continue with the derivation of a kinetic description of the system. Finally, we will discuss simulation results with a particular focus on the emergence of large-scale collective motion. 

\section{Individual Based Model}

We consider a system of $N$ self-propelled particles in two spatial dimensions, which move with a constant speed $s_0$ in a spatial domain of size $L\times L$ with periodic boundary conditions. The interaction between different particles (individuals) is modelled as an effective social force ${\bf F}_i$. 
The evolution of the system is determined by the following equations of motion for the positions ${\bf r}_i$  and the polar orientation angle $\varphi$, which determines the direction of the heading unit vector ${\bf e}_{h,i}(t)$:
\begin{align}
\dot {\bf r}_i & = s_0 {\bf e}_{h_i}(t) = s_0 \left( \begin{array}{c} \cos\varphi_i(t) \\ \sin\varphi_i(t)  \\
\end{array}\right),\label{eq:eom_pos} \\ 
\dot \varphi_i & = \frac{1}{s_0}\left( F_{i,\varphi} + \sqrt{2 D_\varphi} \xi_\varphi\right) \label{eq:eom_angle}\,
\end{align}

The temporal evolution of $\varphi_i$ is determined by the turning of the individual due to social interactions $F_{i,\varphi}$ and random (angular) fluctuations with the intensity $D_\varphi$.    
The (angular) social force is given by the projection of the total social force vector $F_{i,\varphi}={\bf F}_i{\bf e}_{\varphi_i}$ on the angular degree of freedom with ${\bf e}_{\varphi_i}=(-\sin\varphi_i,\cos\varphi_i)$.  
The angular noise $\xi_\varphi$ is Gaussian white noise with zero mean and vanishing temporal correlations.

The total social force is given by a sum of three components:
\begin{align}
{\bf F}_i={\bf f}_r+{\bf f}_m +{\bf f}_a .
\end{align}
The first term represents a short range repulsion responsible for collision avoidance. It reads 
\begin{align}\label{eq:epspp_repuls}
{\bf f}_r & = -\frac{\mu_r}{N_r} \sum_{j=1}^N \hat {\bf r}_{ji}\theta(l_r- r_{ji}),
\end{align}
with $\mu_r \geq 0 $ being a constant repulsive turning rate. The Heaviside function $\theta(l_r- r_{ji})$ ensures that the repulsion takes place only if the distance $r_{ji}=|{\bf r}_j-{\bf r}_i|$ between the focal individual $i$ and the respective neighbor $j$ is below the repulsion distance $l_r$. The total repulsive response is normalized by the number of individuals within the repulsion distance 
\begin{align}
N_r=N_r(t)=\sum_{i=1}^N \theta(l_r-r_{ji}). 
\end{align}
The other two forces read:
 \begin{align}
 {\bf f}_m & = \frac{1}{N_m(t)} \sum_{j=1}^N \mu_a  |\tilde v_{ji}| \hat {\bf r}_{ji} 
 \theta(l_s -  r_{ji})\theta( r_{ji}-l_{r})\theta(+\tilde v_{ji}), \label{eq:fm}\\
 {\bf f}_a & = \frac{1}{N_a(t)} \sum_{j=1}^N \mu_m  |\tilde v_{ji}| \hat {\bf r}_{ji}  
 \theta(l_s -  r_{ji})\theta( r_{ji}-l_{r})\theta(-\tilde v_{ji}). \label{eq:fa}  
 \end{align}
Both forces represent averaged two-individual interactions, which act always along the unit vector pointing towards the center of mass of the neighboring particle $\hat {\bf r}_{ji}=({\bf r}_j-{\bf r}_i)/|{\bf r}_j-{\bf r}_i|$.
 The first one, ${\bf f}_m$, represents the response to approaching individuals characterized by a negative relative velocity $\tilde v_{ji}=({\bf v}_j-{\bf v}_i){\bf \hat r}_{ji}<0$.   
 The second, ${\bf f}_m$, is the corresponding response to moving away (or receding) individuals characterized by positive relative velocity $\tilde v_{ji}>0$. This differentiation is reflected by the last Heaviside functions $\theta(\pm \tilde v_{ji})$. The two other step functions are identical for both interactions and restrict these social responses to neighbors within a sensory range $l_s$ but outside the repulsion zone. The parameters $\mu_{m,a}$ determine the turning rates due to the respective interaction.
Both force terms are proportional to the relative velocity, which lead to stronger responses to faster approaching or receding individuals. Furthermore, they are normalized by the respective number of individuals for the corresponding interaction type: 
\begin{align}
N_m(t) & =\sum_{i=1}^N \theta(r_{ji}-l_r)\theta(l_s-r_{ji})\theta(+\tilde v_{ji}), \quad N_a(t)=\sum_{i=1}^N \theta(r_{ji}-l_r)\theta(l_s-r_{ji})\theta(- \tilde v_{ji}). 
\end{align}

Here, we used for simplicity step-like functions for the spatial dependence of the different interaction. The general results will not be altered by other smooth functions of the distance as long as they decay sufficiently fast in order to ensure local interactions. Please note that the definition of the step-like interaction zones resembles the two-zone model introduced by Couzin and co-workers \cite{couzin_effective_2005,couzin_uninformed_2011}. However, the model discussed here does not contain an explicit velocity-alignment.  

\begin{figure}
\begin{center}
\centering\includegraphics[width=0.5\linewidth]{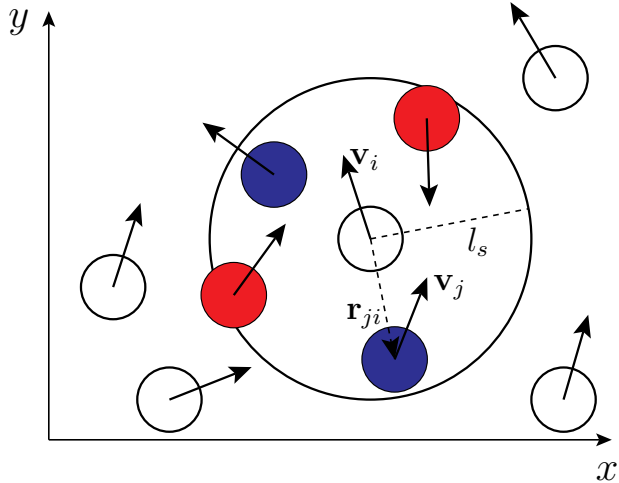}
\caption{Schematic visualization of social interactions: The focal individual $i$ (gray) can interact with individuals within its sensory range $l_s$. Hereby it distinguishes between ``approaching'' (red) and ``moving away'' (blue) individuals.  The decisive factor in the distinction is the sign of relative velocity $\tilde v_{ji}$ defined by the projection of the velocity difference  of neighbor $j$ and the focal individual (${\bf v}_{ji}={\bf v}_j-{\bf v}_i$) on the relative position unit vector ${\bf \hat r}_{ji}={\bf r}_{ji}/|{\bf r}_{ji}|$.     \label{fig:scheme}}
\end{center}
\end{figure}

A schematic visualization of the interaction scheme with the differentiation between approach and moving away is given in Fig \ref{fig:scheme}.

 %Recently, it was argued based on empirical data that interaction of starling within a flock is based rather on a topological than a metric distance, so that a bird interacts only with a finite number of its neighbors. In reality,    
 %Please note that for $\mu_a\leq0$ and $\mu_m\geq0$ the response terms are similar to those of the original model as defined in Eq. \eqref{eq:ep_respS}. In contrast to the original model, we do not formulate any restrictions on the sign of the response strengths  $\mu_{m,a}$ ($\mu_{m,a} \in \mathbb{R}$), which leads to different dynamical regimes, discussed further below. 
%We restrict here to the cases where the repulsion turning rate $\mu_r$ is larger then the maximal turning rate for the other two interaction types:
%\begin{align}
%\mu_r>\text{Max}\left(\mu_m|v_\text{rel}|,\mu_a|v_\text{rel}|\right)=2 v_0 \text{Max}(\mu_m,\mu_a)
%\end{align} 

\begin{figure}
\begin{center}
\centering\includegraphics[width=\linewidth]{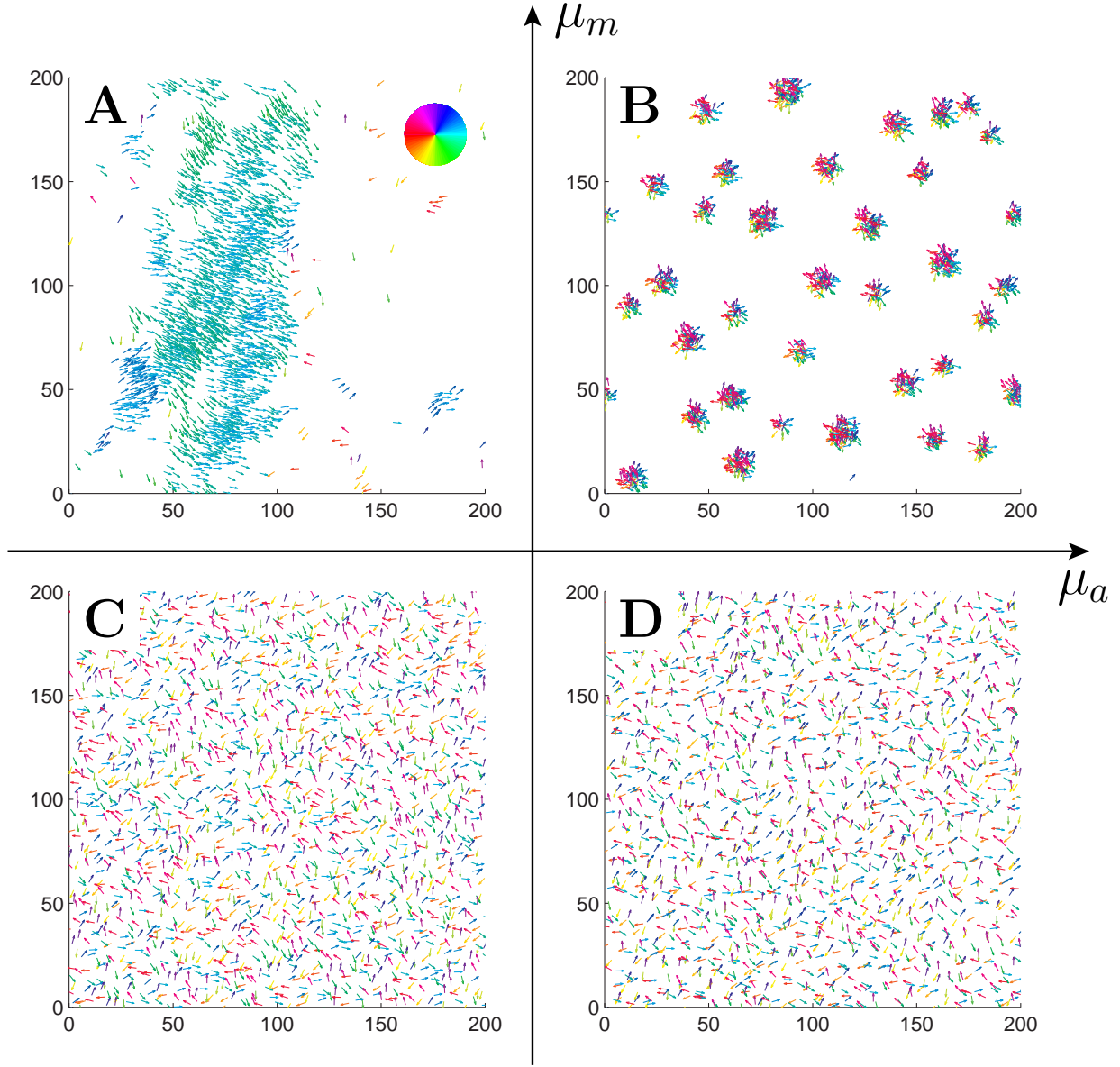}
\caption{Examples of spatial configurations for different regimes: (A) ``Escape and Pursuit'' ($\mu_a=-3.0$, $\mu_m=3.0$), (B) ``Pure Attraction'' ($\mu_a=3.0$, $\mu_m=3.0$), (C) ``Pure Repulsion'' ($\mu_a=-3.0$, $\mu_m=-3.0$) and (D) ``Head on Head'' ($\mu_a=+3.0$, $\mu_m=-3.0$). The arrows and their color indicate the direction of motion of individual particles. The mapping of the color to the directions is shown as an inset in (A). The different panels are arranged according to the location of the corresponding regime in the interaction parameter space, with the origin ($\mu_m=\mu_a=0$) being in the center.    \label{fig:epspp_snapshots}}
\end{center}
\end{figure}

The social forces ${\bf f}_{a/m}$ can lead independently to a repulsive (attractive) response to approaching individuals for $\mu_a<0$ ($\mu_a>0$) and a repulsion (attraction) to individuals moving away $\mu_m<0$ ($\mu_m>0$).
In the $\mu_m\mu_a$-parameter space we distinguish the four quadrants corresponding to different behavior types (see also Fig. \ref{fig:epspp_snapshots}):
\renewcommand{\theenumi}{\roman{enumi}}
\begin{enumerate}
\item {\bf Pure Repulsion:}     repulsion from approaching and moving away individuals: $\mu_a < 0$ and $\mu_m < 0$.   
\item {\bf Escape and Pursuit:} repulsion from approaching individuals  $\mu_a < 0$, attraction to moving away individuals  $\mu_m > 0$.   
\item {\bf ``Head on Head'':}   attraction to approaching individuals $\mu_a > 0$, repulsion from moving away individuals $\mu_m < 0$.   
\item {\bf Pure Attraction:}    attraction to approaching and moving away individuals: $\mu_a > 0$ and $\mu_m > 0$.   
\end{enumerate}
There exist also the special cases with $\mu_{m/a}=0$ and $\mu_{a/m}>0$ ($\mu_{a/m}<0$), which correspond to a selective attraction (repulsion) only to approaching/moving away individuals, and the case of particles interacting only via short-range repulsion ($\mu_{m}=\mu_{a}=0$).

We refer to the situation $\mu_a<0$ and $\mu_m>0$ as ``Escape and Pursuit'', due similar behavior as in the original Brownian particle model \cite{romanczuk_collective_2009}. For $\mu_a>0$ and $\mu_m<0$ the social forces lead to a preference to move towards other individuals which already are coming closer and therefore favor (in particular at low densities) frontal collisions between individuals. We refer to this regime as ``Head on Head''.

%%%%%%%%%%%%%%%%%%%%%%%%%%%%%%%%%%%%%%%%%%%%%%%%%%%%%%%%%%%%%%%%%%%%%%%%%%%%%%%%%%%%%%%%%%%%%
%%%%%%%%%%%%%%%%%%%% BEGIN KINETIC THEORY %%%%%%%%%%%%%%%%%%%%%%%%%%%%%%%%%%%%%%%%%%%%%%%%%%%
%%%%%%%%%%%%%%%%%%%%%%%%%%%%%%%%%%%%%%%%%%%%%%%%%%%%%%%%%%%%%%%%%%%%%%%%%%%%%%%%%%%%%%%%%%%%%

\section{Kinetic Description}
\label{sec:kinetic}

In this section we derive a kinetic description for the above individual based model. 
For this purpose we introduce the $N$-particle probability density function (PDF)
$$P_N({\bf r}_1,\varphi_1; {\bf r}_2,\varphi_2; \ldots ;
{\bf r}_N,\varphi_N ; t)\ ,$$ 
which determines the probability to find a particle (individual) at time $t$, at position ${\bf r}_i$, 
with velocity pointing in direction $\varphi_i$ ($i=1, 2, \ldots, N$). It is normalized with
respect to integration over space and over all angles. Further on, for simplicity, we assume that correlations between particles can be
neglected. Therefore, the $N$-particle distribution density shall
factorize, i.e.  $P_N=\Pi_{i=1}^N P({\bf r}_i,\varphi_i,t)$. In
agreement with \eqref{eq:eom_pos} and \eqref{eq:eom_angle}, we can write down the Fokker-Planck equation (FPE) for the PDF of the $i$-th particle
\begin{align}
  \label{eq:fpe1}
    \frac{\partial}{\partial t} \,P({\bf r}_i,\varphi_i,t)\, =
\,-\, \frac{\partial}{\partial {\bf r}_i} {\bf e}_{h_i}\,P \,
-\,\frac{1}{s_0} \frac{\partial}{\partial \varphi_i} {\bf F}_i {\bf e}_{\varphi_i} P\, 
+\,\frac{D_{\varphi}}{s_0^2} \frac{\partial^2}{\partial \varphi_i^2}\, P\,,
\end{align}
with ${\bf e}_{h_i}=( \cos\varphi_i, \sin\varphi_i)^T$ being the unit vector in the heading direction of individual $i$, and ${\bf e}_{\varphi_i}=( -\sin\varphi_i, \cos\varphi_i)^T$ being the angular unit vector perpendicular to ${\bf e}_{h_i}$.
The above FPE is nonlinear since the interaction force ${\bf F}_i$  depends on the probability
density for the position and the velocity angle of the particles within
their sensory range.

We will now reduce the description to moments of the one-particle PDF,
which are the particle density $\rho({\bf r},t)$
\begin{eqnarray}
  \label{eq:density}
  \rho({\bf r},t)\,=\, \int_0^{2\pi} {\rm d}{\bf r}\, P({\bf r},\varphi,t)\,,~~~~\int_V {\rm d}{\bf r}\,\rho({\bf r},t)\,=\,1\,, 
\end{eqnarray}
and the expectation values of the cosine and sine of the velocity
angle defined as
\begin{eqnarray}
  \label{eq:mean_trig}
  s({\bf r}_i,t)\,=\,\int_0^{2\pi} {\rm d}\varphi \sin(\varphi) P(\varphi,t|{\bf r}_i)\,,~~~~~~~~c({\bf r}_i,t)\,=\,\int_0^{2\pi} {\rm d}\varphi \cos(\varphi) P(\varphi,t|{\bf r}_i)\,.
\end{eqnarray}
where the conditional probability density function of the velocity
direction $\varphi$ is defined through the relation $P({\bf r},\varphi,t)=P(\varphi,t|{\bf r}_i)
\rho({\bf r},t)$. Similar approach was previously used in the context
of swarming of Active Brownian Particles in
\cite{romanczuk_collective_2010,romanczuk_mean-field_2011,grossmann_active_2012}.

Integrating the Fokker-Planck equation \eqref{eq:fpe1} over the velocity-angle yields the continuity equation which reads
\begin{eqnarray}
  \label{eq:cont}
  \frac{\partial}{\partial t}\,\rho({\bf r}_i,t)\,=\,-\, s_0\frac{\partial}{\partial x_i}\,c({\bf r}_i,t)\,\rho({\bf r}_i,t) \,-\,s_0\frac{\partial}{\partial y_i}\,s({\bf r}_i,t)\,\rho({\bf r}_i,t)\,.
\end{eqnarray}
Similarly, we derive the equations for the angular moments
\begin{eqnarray}
  \label{eq:ang}
  \frac{\partial}{\partial t}\,c({\bf r}_i,t)\rho({\bf r}_i,t)\,&=&\,-\, s_0\frac{\partial}{\partial x_i}\,c_2({\bf r}_i,t)\,\rho({\bf r}_i,t) \,-\,s_0\frac{\partial}{\partial y_i}\,cs({\bf r}_i,t)\,\rho({\bf r}_i,t)\,\nonumber \\ 
  && -\,\frac{D_{\varphi}}{s_0^2} c({\bf r}_i,t) \rho({\bf r}_i,t) \,-\,\frac{1}{s_0}\,\langle{\bf F}_i {\bf e}_{\varphi_i} \sin\varphi_i\rangle\,\rho({\bf r}_i,t)  \\
  % \,-\,\frac{1}{s_0}\int_0^{2\pi} {\rm d }\varphi_i \,\sin(\varphi_i) \,  {\bf F}_i {\bf e}_{\varphi_i}\,P({\bf r}_i,\varphi_i,t)\nonumber \\
  \frac{\partial}{\partial t}\,s({\bf r}_i,t)\rho({\bf r}_i,t)\,&=&\,-\, s_0\frac{\partial}{\partial x_i}\,cs({\bf r}_i,t)\,\rho({\bf r}_i,t) \,-\,s_0\frac{\partial}{\partial y_i}\,s_2({\bf r}_i,t)\,\rho({\bf r}_i,t)\nonumber \\ 
  && -\,\frac{D_{\varphi}}{s_0^2} c({\bf r}_i,t) \rho({\bf r}_i,t)\,+\,\frac{1}{s_0}\,\langle{\bf F}_i {\bf e}_{\varphi_i} \cos\varphi_i\rangle\,\rho({\bf r}_i,t) 
  % \frac{1}{s_0}\int_0^{2\pi} {\rm d }\varphi_i \,\cos(\varphi_i) \,
  % {\bf F}_i {\bf e}_{\varphi_i}\,P({\bf r}_i,\varphi_i,t)\,.
  % \nonumber
\end{eqnarray}
  Here, we introduced the averaging over the conditional
  PDF as $\langle...\rangle\,=\,\int_0^{2\pi}{\rm d} \varphi\,... \,
  P(\varphi,t|{\bf r})$ which are functions of the position ${\bf r}$
  and time. The second moments of the trigonometric functions of
  the direction angle read
  \begin{eqnarray}
    \label{eq:2ndmoment}
    c_2({\bf r}_i,t)&=&\,\langle\cos^2(\varphi_i)\rangle\,=\,\frac{1}{2}\big(1\,+\,\langle\cos(2\varphi_i)\rangle\big)\, \approx \,\frac{1}{2}\,, \nonumber\\
    s_2({\bf r}_i,t)&=&\,\langle\sin^2(\varphi_i)\rangle\,=\,\frac{1}{2}\big(1-\langle\cos(2\varphi_i)\rangle \big)\,\approx \,\frac{1}{2}\,,\nonumber\\
    cs_({\bf r}_i,t)&=&\,\langle\cos(\varphi_i)\sin(\varphi_i)\rangle\,=\,\frac{1}{2}\,\langle \sin(2\varphi_i)\rangle \,\approx \,0\,.
  \end{eqnarray}
The approximation for the second moments indicated on the r.h.s. of
the above equation, will be used later on. It neglects higher
harmonics in the equation of motion for the angular PDF, which
essentially restricts the analysis to the relaxation of the slowest
(fundamental) modes of the angular PDF.

In order to obtain an equation for the density, we insert the
equations of the angular moments into the continuity equation. For
this purpose, we take the temporal derivative of the spatial density
for a second time 
\begin{eqnarray}
  \label{eq:tele}
  &&\left(\frac{\partial^2}{\partial t^2}\,+\, \frac{D_\varphi}{s_0^2} \frac{\partial}{\partial t} \,\right)\,\rho({\bf r}_i,t)\,= \\
  &&~~~~~~~~~~~~\frac{s_0^2}{2}\left(\frac{\partial^2}{ \partial x_i^2} c_2({\bf r}_i,t) \rho({\bf r}_i,t)\,+\,
    2\frac{\partial^2}{\partial x_i \partial y_i} cs({\bf r}_i,t) \rho({\bf r}_i,t)\,+\,\frac{\partial^2}{ \partial y_i^2} s_2({\bf r}_i,t) \rho({\bf r}_i,t)\right)\nonumber\\
  &&~~~~~~~~~~~~~~~~~~~~~~~~~~+\,\frac{1}{s_0}\,\frac{\partial}{ \partial x_i }\,\langle{\bf F}_i {\bf e}_{\varphi_i} \sin\varphi_i\rangle\,\rho({\bf r}_i,t) -\,\frac{1}{s_0}\,
  \frac{\partial}{ \partial y_i }\,\langle {\bf F}_i {\bf e}_{\varphi_i} \cos\varphi_i\rangle\rho({\bf r}_i,t)  \nonumber
\end{eqnarray}
If we introduce in this equations abbreviations for the variances of
the second angular moments defined as
\begin{eqnarray}
  \label{eq:temp}
&&T_{x_i,x_i}({\bf r}_i,t)\,=\, c_2({\bf r}_i,t)-c^2({\bf r}_i,t)\,,\nonumber\\
&&T_{x_i,y_i}({\bf r}_i,t)\,=\, cs({\bf r}_i,t)-c({\bf r}_i,t)\,s({\bf r}_i,t)\,,\nonumber\\
&&T_{y_i,y_i}({\bf r}_i,t)\,=\, s_2({\bf r}_i,t)-s^2({\bf r}_i,t)\,,
\end{eqnarray}
the equation for the density becomes
\begin{eqnarray}
  \label{eq:density2}
  &&\left(
    \frac{\partial^2}{\partial t^2}\,+\, 
    \frac{D_\varphi}{s_0^2} \frac{\partial}{\partial t} \,\right)
  \,\rho\,= \,\frac{s_0^2}{2}\left(\frac{\partial^2}{ \partial x_i^2}
    \left(T_{x_i,x_i} +c^2 \right) \rho
    \,+\,2\frac{\partial^2}{\partial x_i \partial y_i}
    \left( T_{x_i,y_i} +c\, s \right) \rho\,
    +\,\frac{\partial^2}{ \partial y_i^2} 
    \left( T_{y_i,y_i} +s^2\right)\rho \right) \nonumber \\
  && ~~~~~~~~~~~~~~~~~~~~~~~~~~+
  \,\frac{1}{s_0}\,\frac{\partial}{ \partial x_i }\,
  \langle{\bf F}_i {\bf e}_{\varphi_i} \sin\varphi_i\rangle\,\rho({\bf r}_i,t)\, -\,\frac{1}{s_0}\,
  \frac{\partial}{ \partial y_i }\,
  \langle {\bf F}_i {\bf e}_{\varphi_i} \cos\varphi_i\rangle\rho({\bf r}_i,t)\,.  
\end{eqnarray}
We note, one still needs equations or expressions for the variances
$T_{u_i.v_i}$ which have to be inserted.

Similarly, we separate the density from the angular moments using the
continuity equation. After a few steps, one obtains the equation of
motion for the mean cosine
\begin{eqnarray}
  \label{eq:cos}
&&\left\{\frac{\partial}{\partial t}+c({\bf r}_i,t)\frac{\partial}{\partial x_i}+s({\bf r}_i,t)\frac{\partial}{\partial y_i}\right\}\,c({\bf r}_i,t)\,= \,-\,\frac{\langle{\bf F}_i {\bf e}_{\varphi_i} \sin\varphi_i\rangle}{s_0}
\,-\,\frac{D_{\varphi}}{s_0^2} c({\bf r}_i,t)\nonumber \\ 
%\int_0^{2\pi} {\rm d }\varphi_i \,\sin(\varphi_i) \,  {\bf F}_i {\bf e}_{\varphi_i}\,P({\bf r}_i,\varphi_i,t) \nonumber \\ 
&&~~~~~~~~~~~~~~~~~~~\,-\, \frac{s_0}{\rho({\bf r}_i,t)}\left\{\frac{\partial}{\partial x_i}\,T_{x_i,x_i}({\bf r}_i,t)\,\rho({\bf r}_i,t) \,+\frac{\partial}{\partial y_i}\,T_{x_i,y_i}({\bf r}_i,t)\,\rho({\bf r}_i,t)\right\}\,,
\end{eqnarray}
and for the mean sine
\begin{eqnarray}
  \label{eq:sin}
&&\left\{\frac{\partial}{\partial t}+c({\bf r}_i,t)\frac{\partial}{\partial x_i}+s({\bf r}_i,t)\frac{\partial}{\partial y_i}\right\}\,s({\bf r}_i,t)\,= \,+\,\frac{\langle{\bf F}_i {\bf e}_{\varphi_i} \cos\varphi_i\rangle}{s_0} \,-\,\frac{D_{\varphi}}{s_0^2} s({\bf r}_i,t) \nonumber \\
%\int_0^{2\pi} {\rm d }\varphi_i \,\cos(\varphi_i) \,  {\bf F}_i {\bf e}_{\varphi_i}\,P({\bf r}_i,\varphi_i,t)\nonumber\\ 
&&~~~~~~~~~~~~~~~~~~\,-\, \frac{s_0}{\rho({\bf r}_i,t)}\left\{\frac{\partial}{\partial x_i}\,T_{x_i,y_i}({\bf r}_i,t)\,\rho({\bf r}_i,t) \,+\frac{\partial}{\partial y_i}\,T_{y_i,y_i}({\bf r}_i,t)\,\rho({\bf r}_i,t)\right\},
\end{eqnarray}
respectively. We multiply the first equation \eqref{eq:cos} by the mean
cosine and the second one \eqref{eq:sin} by the mean sine and obtain the equation for the order
parameter:
\begin{eqnarray}
\label{eq:order}
  &&\frac{1}{2}\left\{\frac{\partial}{\partial t}+c({\bf r}_i,t)\frac{\partial}{\partial x_i}+s({\bf r}_i,t)\frac{\partial}{\partial y_i}\right\}\,(c^2({\bf r}_i,t)+s^2({\bf r}_i,t))\,=\,-\,\frac{D_{\varphi}}{s_0^2} (c^2({\bf r}_i,t)+s^2({\bf r}_i,t))\, \nonumber \\ 
&&~~~~~~-\,\frac{1}{s_0}\,\big(\,c({\bf r}_i,t)\,\langle\sin(\varphi_i) {\bf F}_i {\bf e}_{\varphi_i}\rangle\,-\,s({\bf r}_i,t)\,\langle\cos(\varphi_i){\bf F}_i {\bf e}_{\varphi_i}\rangle\,\big)  \nonumber\\ 
&&~~~~~~~~~~~~~~~\,-\, \frac{s_0\,c({\bf r}_i,t)}{\rho({\bf r}_i,t)}\left\{\frac{\partial}{\partial x_i}\,T_{x_i,x_i}({\bf r}_i,t)\,\rho({\bf r}_i,t) \,+\frac{\partial}{\partial y_i}\,T_{x_i,y_i}({\bf r}_i,t)\,\rho({\bf r}_i,t)\right\}\\ 
&&~~~~~~~~~~~~~~~~~~~~~~\,-\, \frac{s_0\,s({\bf r}_i,t)}{\rho({\bf r}_i,t)}\left\{\frac{\partial}{\partial x_i}\,T_{x_i,y_i}({\bf r}_i,t)\,\rho({\bf r}_i,t) \,+\frac{\partial}{\partial y_i}\,T_{y_i,y_i}({\bf r}_i,t)\,\rho({\bf r}_i,t)\right\}\,.\nonumber
\end{eqnarray}
The only term, which can induce an instability of the homogeneous,
disordered solution in Eqs. (\ref{eq:tele}) and (\ref{eq:order})
contains the interaction force. The remaining terms, which contain
only the zeroth up to second moments of the orientation, are present
also in the case of non-interacting particles. They describe the
relaxation dynamics of the resulting patterns towards a steady state,
but are not the source of an eventual inhomogeneous solutions.

We see that a further analysis needs still the treatment of integrals
which contain the interaction forces, which can be very tedious in the
general case. Further below, we will consider principal cases where
$\mu_m$ and $\mu_a$ have the same absolute value and differ only in
their sign and derive approximations for the respective integrals.

The force changing the direction of the $i$th-particle can be
formulated by introducing an interaction parameter depending of the
distance between particles and their relative velocities. We define
\begin{equation}
	\mu(r_{ji},{v}_{ji})\,=\, 
	\begin{cases}
          -  \frac{\mu_r}{N_r}     & \text{ for $0 \le r_{ji}  \le l_r$} \\
              \mu_z(\tilde v_{ji}) & \text{ for $l_r < r_{ji} \le l_s$}   	
	\end{cases}\,, 
	\label{mu1}
\end{equation}
where $r_{ji}$ is the distance between from the $i$th to the $j$th
particle and $\tilde v_{ji}$ again the relative velocity projected on
the distance vector. Otherwise, this function vanishes.
The $\mu_z$ have to be specified for positive and negative
 relative velocities. They read, in agreement with \eqref{eq:fm} and
\eqref{eq:fa},
\begin{equation}
  \mu_z({v}_{ji})\,=\, 
  \begin{cases}
    -  \frac{\mu_a \, \tilde v_{ji}}{N_a}  & \text{ for $\tilde v_{ji}  \le 0$} \\
    + \frac{\mu_m \, \tilde v_{ji}}{N_m} & \text{ for $\tilde v_{ji} >
      0 $}
  \end{cases} \,.
  \label{mu2}
\end{equation}
The normalizing numbers in denominators can be expressed as integrals
over the respective distances and relative velocities. In detail, with the definition of the total number density as
$n_0=N/V$, where $V=L^2$ is the total volume (area in 2d) one gets
\begin{align}
  \label{eq:norm}
  N_z\,= \frac{\Delta V_z}{V}\,N\,=\,n_0\int_{\Delta V_z}{\rm d}{\bf
    r}_{ji}\int_0^{2\pi}{\rm d}\varphi_j P(\varphi_j,t|{{\bf r}+\bf
    r}_{ji})\,\rho({{\bf r}+\bf r}_{ji},t)\,,
\end{align}
where the ${\Delta V_z}, z=r,a,m$ are the (effective) partial volumes for the different
interaction types, respectively. For the integration one can
transform to polar coordinates, integrating over the distance $r_{ji}$
and the polar angle $\alpha_{j}$, determining the position of
particle $j$ with respect to the focal particle $i$. 

The force in the third term on the r.h.s. of the Fokker-Planck equation (FPE, Eq. \ref{eq:fpe1}) can be expressed as an integral over the probability density of the independent other particles. With the new definition of the interaction parameter, this forces reads 
\begin{eqnarray}
  \label{eq:force1}
  {\bf F}_i\,=\, \sum_{i=1}^N \hat {\bf r}_{ji} \,\mu(|{\bf r}|_{ji}, \tilde v_{ji})\,.
\end{eqnarray}
Expressed by the probability density of the $N$ interacting particles
it becomes
\begin{eqnarray}
  \label{eq:force2}
  {\bf F}_i\,=\, n_0 \int_V{\rm d}{\bf r}_{ji} 
  \langle\hat {\bf r}_{ji}  \,\mu(|{\bf r}|_{ji}, \tilde v_{ji})\rangle_{\varphi_j} \,\rho({\bf r}_i+{\bf r}_{ji},t)\,.
\end{eqnarray}

Finally, we derive an expression of the projection of the relative velocity. After a few calculations, one obtains the relative velocity as a function of the velocity angles $\varphi_{i}$, $\varphi_{j}$ and the position angle $\alpha_j$:
\begin{eqnarray}
  \label{eq:velo}
  \tilde v_{ji} = \tilde v_{ji}(\varphi_i,\varphi_j,\alpha_j)=\,- 2 {s}_0 \sin\left(\frac{\varphi_j+\varphi_i}{2}-\alpha_j\right)\,\sin\left(\frac{\varphi_j-\varphi_i}{2}\right)\,,
\end{eqnarray}
which allows a closure of the kinetic description.

By virtue of the assumptions made, the effective force separates in three
parts: one coming from the repulsion at short distances, one from
``moving away'' and the third, from ``approaching'' individuals. In the following discussion, we will first restrict ourselves to the second and the third interaction type, which corresponds to a limit of a vanishing short-range repulsion length $l_r=0$. In general, the short-ranged repulsion increases the pressure and the effective temperature in the
system but does not induce instabilities and inhomogeneous steady states as observed in the
simulations (see Sect \ref{sec:simulation}). However, it determines the size of the observed
structures. 

In the following, we investigate the force turning the velocity angle and
project on the unit vector ${\bf e}_{\varphi_i}(t)$. The integral (\ref{eq:force2})
transforms into
\begin{align}
  \label{eq:force3}
  {\bf F}_i {\bf e}_{\varphi_i}\,=\,n_0 \,{\bf e}_{\varphi_i}\Big\{ & 
  \frac{\mu_m}{N_m}\,\int_{r_{ji}<l_s} {\rm d}{\bf r}_{ji}\,\langle \tilde v_{ji} \theta(+\tilde v_{ji})
  \hat {\bf r}_{ji}\rangle_{ \varphi_j}\, \rho({\bf r_i}+{\bf r}_{ji},t)\,- \nonumber \\
   &-\,\frac{\mu_a}{N_a}\,\int_{r_{ji}<l_s}{\rm d}{\bf r}_{ji} 
 \,\langle \tilde v_{ji} \theta(-\tilde v_{ji})
  \hat {\bf r}_{ji}\rangle_{ \varphi_j}\, \rho({\bf r_i}+{\bf r}_{ji},t)\,\Big\}
\end{align}
where the spatial distance between the $j$-th and the $i$-th particle
appears explicitly in the argument of the PDF.  The product between
the two unit vectors in (\ref{eq:force3}) can be expressed in terms of
the different angles as ${\bf e}_\varphi \cdot \hat {\bf
  r}_{ji}=\sin(\alpha_j-\phi_i)$.  Further on, we will also not
distinguish from the particle number approaching and moving away by
setting $N_m=N_a$.

\begin{figure}
\begin{center}
\centering\includegraphics[width=0.5\linewidth]{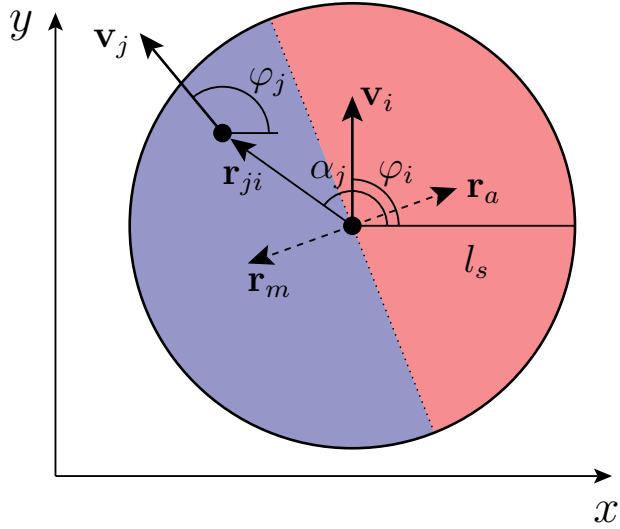}
\caption{Visualization of the support in the spatial integration for ``approach'' and ``moving away'' in the kinetic description for a binary interaction of the $i$-th individual (particle), moving with velocity ${\bf v}_i$ (heading angle $\varphi_i$), with a neighbor $j$ within its sensory range $l_s$ moving with velocity ${\bf v}_j$ (heading angle $\varphi_j$). The dotted line, determined by the mean angle $(\varphi_i+\varphi_j)/2$, represents the border between the two distinct spatial regions (half-spheres) corresponding to ``approach'' and ``moving away'' of individual $j$: If the relative position vector ${\bf r}_{ji}$ of individual $j$ with angle $\alpha_j$ points above the dotted line, into the red half-sphere, then the two particles are coming closer (``approach''). If ${\bf r}_{ji}$ points below the dotted line (blue half-sphere) then the two particles move away from each other (as in this example). The dashed vectors correspond to the center-of mass of the correspond
 ing half-spheres.
\label{fig:scheme_angles}}
\end{center}
\end{figure}

From Eq. \eqref{eq:velo} we find the two different regions of
integration where the relative velocity has a different sign for $\varphi_j$ in the interval $\varphi_i\le\varphi_j\le 2\pi+\varphi_i$:
\begin{equation}
  \begin{array}{c c r c l c}
    v_{ji}  > 0   & \text{for} & \frac{\varphi_j+\varphi_i}{2}     & < \alpha_j \le & \frac{\varphi_j+\varphi_i}{2}+\pi  & \text{``moving away'',}\\
    v_{ji}  < 0   & \text{for} & \frac{\varphi_j+\varphi_i}{2}+\pi & < \alpha_j \le & \frac{\varphi_j+\varphi_i}{2}+2\pi & \text{``approaching'',}
  \end{array} 
  \label{vrel}
\end{equation}
Hence, a particle, located in the half-sphere
in clockwise direction from the mean angle
$(\varphi_j+\varphi_i)/2$ approaches the focal particle, whereas,
a particle located in the other half-sphere counterclockwise from the
mean angle are moving away. Thus, the support of the interaction integrals for ``approach'' and ``moving away'' corresponds to these two different half-spheres (see Fig \ref{fig:scheme_angles}). Please note that for $\varphi_j=\varphi_i$ the social force vanishes as $\tilde v_{ji}=0$ (\ref{eq:velo}). 

As an approximation, we will replace this integration by
averaged values fixing the probability distribution in the centers of
the half-spheres, i.e. at a distance from $l_s/2$ and with the angle
$\alpha_{j,a}=(\varphi_j+\varphi_i)/2 -\pi/2$ for the approaching particles and
correspondingly $\alpha_{j,m}=(\varphi_j+\varphi_i)/2 +\pi/2$ for particles moving
away. The corresponding distance vectors are parallel but point in a
different direction. They  read
\begin{eqnarray}
  \label{eq:distances}
  {\bf r}_m\,=\,\frac{l_s}{2}\, {\bf e}_{\frac{\varphi_j+\varphi_i +\pi}{2}}\,,~~~~~~~~~~{\bf r}_a\,=\,-\frac{l_s}{2} \, {\bf e}_{\frac{\varphi_j+\varphi_i +\pi}{2}}\,.
\end{eqnarray}
These vectors correspond to the center of mass of approaching (moving away) individuals under the assumption of a homogeneous spatial distribution of neighbors with direction $\varphi_j$.

We separate the spatial PDF from the distribution of the
velocity-angles using the conditional angle PDF, i.e. $P({\bf r},\varphi,t)=\rho({\bf r},t)
P(\varphi,t|{\bf r})$. Assuming that the density depends only weakly
on the position within the sensing region, one can take it
constant within the range of spatial integration. Consequently, the
spatial integration cancels together with the denominator and we get for
the projected force
\begin{eqnarray}
  \label{eq:force4}
  {\bf F}_i {\bf e}_{\varphi_i}\,&=&+\,s_0 \,\mu_m\, \int_{\varphi_i}^{\varphi_i+ 2\pi} {\rm d}\varphi_j \sin(\varphi_j-\varphi_i)\, P(\varphi_j,t|{\bf r}_i+{\bf r_m})\,\\
  &&~~~~~~~~	-\,s_0 \,\mu_a\,\, \int_{\varphi_i}^{\varphi_i+ 2\pi} {\rm d}\varphi_j \sin(\varphi_j-\varphi_i)\, P(\varphi_j,t|{\bf r}_i+{\bf r_a}\,)  \nonumber 
\end{eqnarray}
%\begin{eqnarray}
%  \label{eq:force4}
%  {\bf F}_i {\bf e}_{\varphi_i}\,= \,s_0 \,\int_{\varphi_i}^{\varphi_i+ 2\pi} {\rm d}\varphi_j \left\{\mu_m \rho({\bf r}_i+{\bf r_m})\,-\,\mu_a \rho({\bf r}_i+{\bf r_a}) \right\} 
%  \cos\left(\frac{\varphi_j-\varphi_i}{2}\right)\, P(\varphi,t|{\bf r}_i))
%\end{eqnarray}
The resulting expression will be taken in a dipole approximation.
Subsequently, we develop the difference in small $l_s$ up to the first
derivative which yields two contributions ${\bf F}_i {\bf
  e}_{\varphi_i}\approx{\bf F}_i {\bf e}_{\varphi_i}^{(1)}+{\bf F}_i
{\bf e}_{\varphi_i}^{(2)}$. In first order, we get 
\begin{align}
  \label{eq:first}
  {\bf F}_i {\bf e}_{\varphi_i}^{(1)}&=\,(\mu_m\,-\,\mu_a) \,\int_{\varphi_i}^{\varphi_i+ 2\pi} {\rm d}\varphi_j \,\sin(\varphi_j-\varphi_i)\, P(\varphi_j,t|{\bf r}_i)\\
  &~~~~~~~~~~~~~~~~~ =\,(\mu_m\,-\,\mu_a)\,\left(\cos(\varphi_i)\,s({\bf r}_i,t) -\sin(\varphi_i)\,c({\bf r}_i,t)\right) \nonumber
\end{align}
The second approximation includes the derivatives in direction of the
centers of the half-spheres. Inserting the corresponding directional
cosines we obtain in second approximation
\begin{eqnarray}
  \label{eq:second}
  {\bf F}_i {\bf e}_{\varphi_i}^{(2)}\,=\,
  \frac{l_s}{2}(\mu_m\,+\,\mu_a)
  \left[ \cos(\varphi_i)\,\langle\,\hat{{\bf r}}_m
  \frac{\partial}{\partial {\bf r}_i}\sin(\varphi_j)\rangle_j \,- 
  \sin(\varphi_i)\,\langle\,\hat{{\bf r}}_m
  \frac{\partial}{\partial {\bf r}_i} \cos(\varphi_j)\rangle_j \right ]
\end{eqnarray}
and the unit vector $\hat{{\bf r}}_m$ of (\ref{eq:distances}) still
depends of the velocity angles which have to be taken into account if
averaging $\langle ... \rangle_j=\int {\rm d} \varphi_j ... P(\varphi_j,t|{\bf r}_i)$.  

Both orders depend qualitatively different of the social force
strengths. Whereas the first order, corresponding to a monopole
contribution, depends on the difference of the interaction strengths,
the second (dipole contribution) contains the sum of both
coefficients.  Assuming that both coefficients differ only in their
sign, the first order describes effectively a perfectly symmetric
``escape and pursuit'' (or ``Head on Head'') situation.  In this case
${\bf F}_i {\bf e}_{\varphi_i}^{(2)}$, which depends on the local gradients, vanishes. 
Thus in this special cases, the effective turning rate does not depend, up to second order, on
spatial (density) inhomogeneities. On the
other hand for social interactions with equal sign, such as ``pure
attraction'' and ``pure repulsion'', the ${\bf F}_i {\bf
  e}_{\varphi_i}^{(1)}$ vanishes and only the second order
contributes. In this case, there is no monopole force driving
the local collective dynamics. However, depending on the sign, the
social force can lead to an amplification of density
inhomogeneities (``pure attraction''), or will stabilize the spatially
homogeneous state (``pure repulsion'').  We should mention that the
short ranged repulsion is described by a similar expression as for the
case of pure repulsion but without the factor $s_0$.

In the general case of arbitrary $\mu_a$ and $\mu_m$, the resulting dynamics will be a combination of both contributions. 
However, from the analysis of the dependence of the effective turning as a function of the interaction strengths, 
it can be easily seen that that the main diagonals $\mu_m=\mu_a$ and $\mu_m=-\mu_a$ separate qualitatively different regions. 
For example, in the escape and pursuit quadrant, the second order contribution changes its sign at the diagonal with $|\mu_a|=|\mu_m|$, 
which indicates that the social interactions tend to amplify density inhomogeneities in the pursuit dominated region $|\mu_m|>|\mu_a|$, 
whereas the opposite is the case in the escape dominated region $|\mu_m|<|\mu_a|$.  

Further on, we will focus for simplicity on the monopole term.  
The averaged force terms in Eqs. (\ref{eq:ang}) become
\begin{align}
  \label{eq:second1}
  &\langle{\bf F}_i {\bf e}_{\varphi_i}^{(1)} \sin\varphi_i\rangle
  \,=\,s_0\,(\mu_m\,-\,\mu_a) \, \rho({\bf
    r}_i,t)\,\left\{(T_{x_i,y_i}({\bf r}_i,t))\,s({\bf
      r}_i,t·) \,-\,T_{y_i,y_i}({\bf r}_i,t)\,c({\bf
      r}_i,t)\right\}\,,
\end{align}
and 
\begin{align}
  \label{eq:second2}
  &\langle{\bf F}_i {\bf e}_{\varphi_i}^{(1)} \cos\varphi_i\rangle
  \,=\,s_0\,(\mu_m\,-\,\mu_a) \, \rho({\bf
    r}_i,t)\,\left\{(T_{x_i,x_i}({\bf r}_i,t))\,s({\bf
      r}_i,t·) \,-\,T_{x_i,y_i}({\bf r}_i,t)\,c({\bf
      r}_i,t)\right\}\,,
\end{align}
respectively. With these expressions, the term in the evolution equation of the order parameter (Eq. \ref{eq:order})
 containing the forces becomes proportional to
$(\mu_m-\mu_a)$ and we obtain
\begin{align}
&-\,\frac{1}{s_0}\,c({\bf r}_i,t)\,\langle\sin(\varphi_i) {\bf F}_i {\bf e}_{\varphi_i}\rangle\,+\,\frac{1}{s_0}\,s({\bf r}_i,t)\,\langle\cos(\varphi_i){\bf F}_i {\bf e}_{\varphi_i}\rangle\,=\\
&~~~~~~(\mu_m\,-\,\mu_a)\,\left\{c^2({\bf r}_i,t)\,T_{x_i,x_i}({\bf r}_i,t)\,+\,s^2({\bf r}_i,t)\,T_{y_i,y_i}({\bf r}_i,t)\,-\,2\,c({\bf r}_i,t)\,s({\bf r}_i,t)\,T_{x_i,y_i}({\bf r}_i,t)\right\}\,.\nonumber
\end{align}
One sees that this term is determined by the temperature of the particle gas. The r.h.s. of the given expression can become positive for positive $\mu_m-\mu_a$. It describes the creation of  order if its value is larger then the effective relaxation process which arises from the noise in the velocity angles. 

For an estimation of the critical values of the onset of order one
might take the approximation in Eq. (\ref{eq:2ndmoment}) which yields
that the r.h.s. approaches $(\mu_m\,-\,\mu_a)\,(c^2({\bf
  r}_i,t)\,+\,s^2({\bf r}_i,t))/2$. Hence, a necessary condition for a growing order parameter, thus onset of collective motion, we obtain 
\begin{align}
\label{crit}
&\mu_m\,-\,\mu_a\,>\, 2 \frac{D_\varphi}{s^2_0} \,.
\end{align}
In general, for vanishing fluctuations, collective motion can emerge only above the main diagonal with $\mu_m>\mu_a$. With increasing noise strength $D_\varphi$ the regime of collective motion, is predicted to recede into the escape-pursuit regime.

%{\color{green}
%We use 
%\begin{eqnarray}
%  \cos\left(\frac{\varphi_j-\varphi_i}{2}\right)\,=\,\sqrt{\frac{1}{2}\left(1+\cos \varphi_j \cos \varphi_i+\sin \varphi_j \sin \varphi_i\right)}
%\end{eqnarray}
%and with decoupling higher moments we obtain 
%\begin{eqnarray}
%  \label{eq:forcefin}
%&&  \langle{\bf F}_i {\bf e}_{\varphi_i}^{(1)} \cos\varphi_i\rangle\,=\,2\,s_0 \,\mu \,\rho^2({\bf r}_i,t)\,c({\bf r}_i,t)\,\sqrt{\frac{1}{2}\left(1+c^2({\bf r}_i,t)+s^2({\bf r}_i,t)\right)}\,,\\
%&&  \langle{\bf F}_i {\bf e}_{\varphi_i}^{(1)} \sin\varphi_i\rangle\,=\,2\,s_0 \,\mu \,\rho^2({\bf r}_i,t)\,s({\bf r}_i,t)\,\sqrt{\frac{1}{2}\left(1+c^2({\bf r}_i,t)+s^2({\bf r}_i,t)\right)}\,.
%\end{eqnarray}
%
%
%In the case of equal sign $\mu_m=\mu_a=\mu$, we obtain for
%the averaged forces terms, using the approximation in Eq.(\ref{eq:2ndmoment}),
%\begin{eqnarray}
%  %\label{eq:second2}
%  \langle{\bf F}_i {\bf e}_{\varphi_i}^{(2)} \cos\varphi_i\rangle\,=\,s_0 \,{l_s}\,\mu_m \,\rho({\bf r}_i,t)) \, \left \{ - \frac{\partial \rho}{\partial x_i}c({\bf r}_i,t) s({\bf r}_i,t) \,+ \frac{\partial \rho}{\partial y_i} \left(\frac{1}{2} + c^2({\bf r}_i,t)\right) \right  \}\,,
%\end{eqnarray}
%and
%\begin{eqnarray}
%  %\label{eq:second1}
%  \langle{\bf F}_i {\bf e}_{\varphi_i}^{(2)} \sin\varphi_i\rangle\,=\,s_0 \,{l_s}\,\mu_m \,\rho({\bf r}_i,t))\,\left\{ \frac{\partial \rho}{\partial x_i}\left(\frac{1}{2}- s^2({\bf r}_i,t)\right) \,+ \frac{\partial \rho}{\partial y_i} c({\bf r}_i,t) s({\bf r}_i,t)\right\}\,,  
%\end{eqnarray}
%respectively.
%}
%
The above result, as well as the formulation of a kinetic description in general, not only provide qualitative insights into the impact of the interaction strengths on the large scale system dynamics, but provide also a starting point for a more quantitative analysis of the stability of the inhomogeneous solutions. However, this requires further assumptions on the properties of the involved probability densities which ensure a closure of the descriptions, e.g. assumptions for the temperature, which are beyond the scope of this work.

%%%%%%%%%%%%%%%%%%%%%%%%%%%%%%%%%%%%%%%%%%%%%%%%%%%%%%%%%%%%%%%%%%%%%%%%%%%%%%%%%%%%%%%%%%%%%
%%%%%%%%%%%%%%%%%%%% END KINETIC THEORY %%%%%%%%%%%%%%%%%%%%%%%%%%%%%%%%%%%%%%%%%%%%%%%%%%%%%
%%%%%%%%%%%%%%%%%%%%%%%%%%%%%%%%%%%%%%%%%%%%%%%%%%%%%%%%%%%%%%%%%%%%%%%%%%%%%%%%%%%%%%%%%%%%%

\section{Simulation Results}
\label{sec:simulation}
In order to characterize the behavior of the model, we have performed systematic numerical simulations for varying interactions strengths ($-5\leq \mu_{a,m} \leq 5$), different densities $\rho$ and noise strengths $D_p$. 

In the following we will discuss our results in term of the dimensionless density $\rho_s=N l_s^2/L^2$, rescaled by the sensory range, which is proportional to the average number of individuals per interaction zone for a homogeneous (random) spatial distribution.

We focus in particular on the question what combinations of $\mu_a$, $\mu_m$ lead to large scale collective motion. The degree of collective motion after the system reaches a steady state is measured using the time averaged center-of-mass speed normalized by the preferred speed of individuals $s_0$, which is the well-known order parameter used in the analysis of Vicsek-type models:
\begin{align}
 \left\langle S \right\rangle_t =  \frac{1}{s_0} \left\langle \left|\langle {\bf v}_{i}\rangle_N\right| \right\rangle_t \ ,
\end{align}
where $\langle \cdot \rangle_t$ denotes the temporal average and $\langle \cdot \rangle_N$ the ensemble average. 
In addition, we measure the spatial inhomogeneity (clustering) by the time averaged scaled neighbor number 
\begin{align}
\langle N \rangle_t = \left\langle  \frac{\langle N_r(t) + N_m(t) + N_m(t) \rangle_N  }{N_\text{max}}  \right\rangle_t,
\end{align}
with $\langle N_r(t) + N_m(t) + N_m(t) \rangle_N$ being the average number of neighbors within the metric distance given by the sensory range $l_s$ of an individual at a given time $t$. The density-dependent scaling number $N_\text{max}$ defines the maximal expectation values for the measured neighbor numbers corresponding to the closest packing of individuals by assuming an impenetrable repulsion zone with a diameter $l_r$:
\begin{align}
N_\text{max}=\eta_{2d}\frac{4  l_s^2}{l_r^2}-1.
\end{align}
Here $\eta_{2d}=\pi/(2\sqrt{3})\approx 0.907$ is the packing fraction for the closest packing of discs in two spatial dimensions. The term $-1$ in the definition of $N_\text{max}$ takes into account that the focal particle is not being counted as its own neighbor. Please note, that as we are considering a soft core interaction, $\langle N \rangle_t$ can in principle be larger than one, in particular for high densities and strong attraction. 

\begin{figure}
\begin{center}
{\large $\rho_s=0.56$, $D_\varphi=0.1$  \vspace{1ex} \\} 
\begin{minipage}{0.47\linewidth}
\centering center of mass speed $\langle S \rangle_t$ \\
\includegraphics[width=0.8\linewidth]{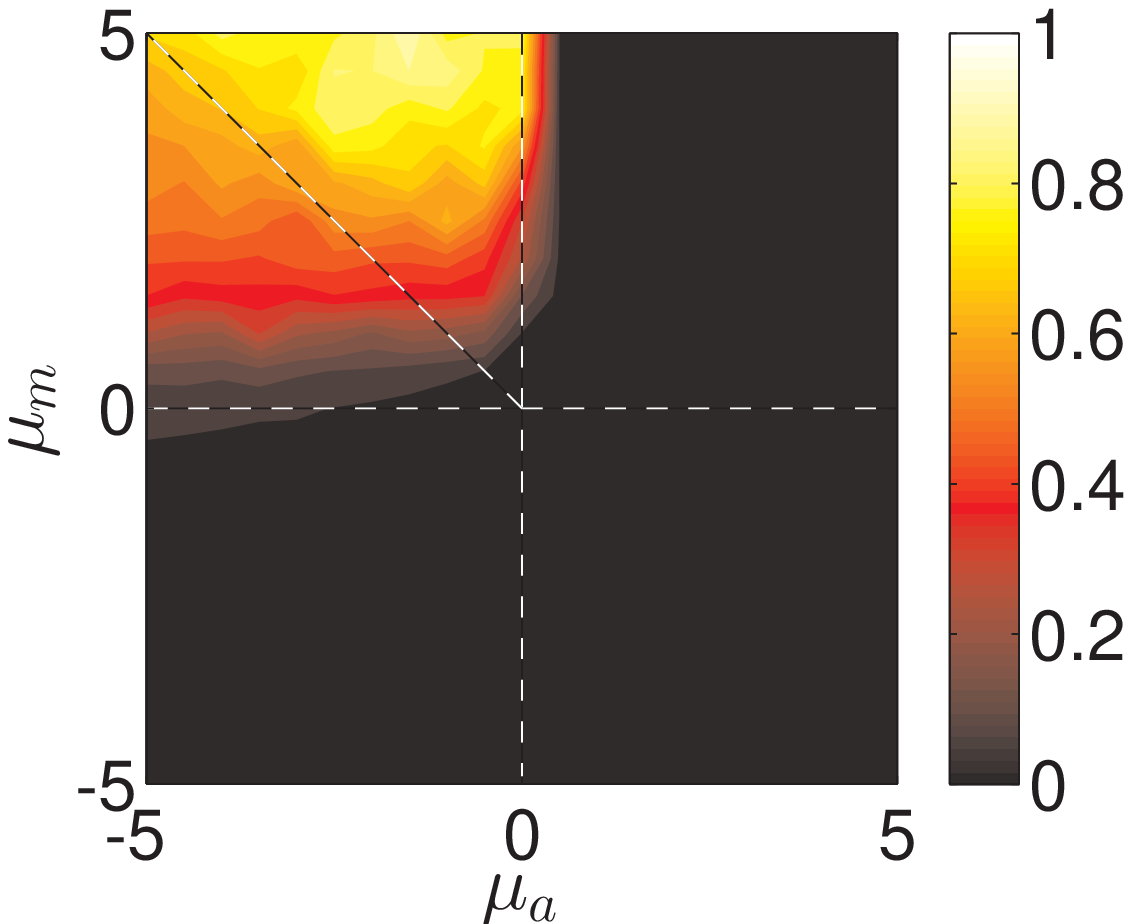}
\end{minipage}
\begin{minipage}{0.47\linewidth}
\centering  neighbor number $\langle N \rangle_t$ \\
\includegraphics[width=0.8\linewidth]{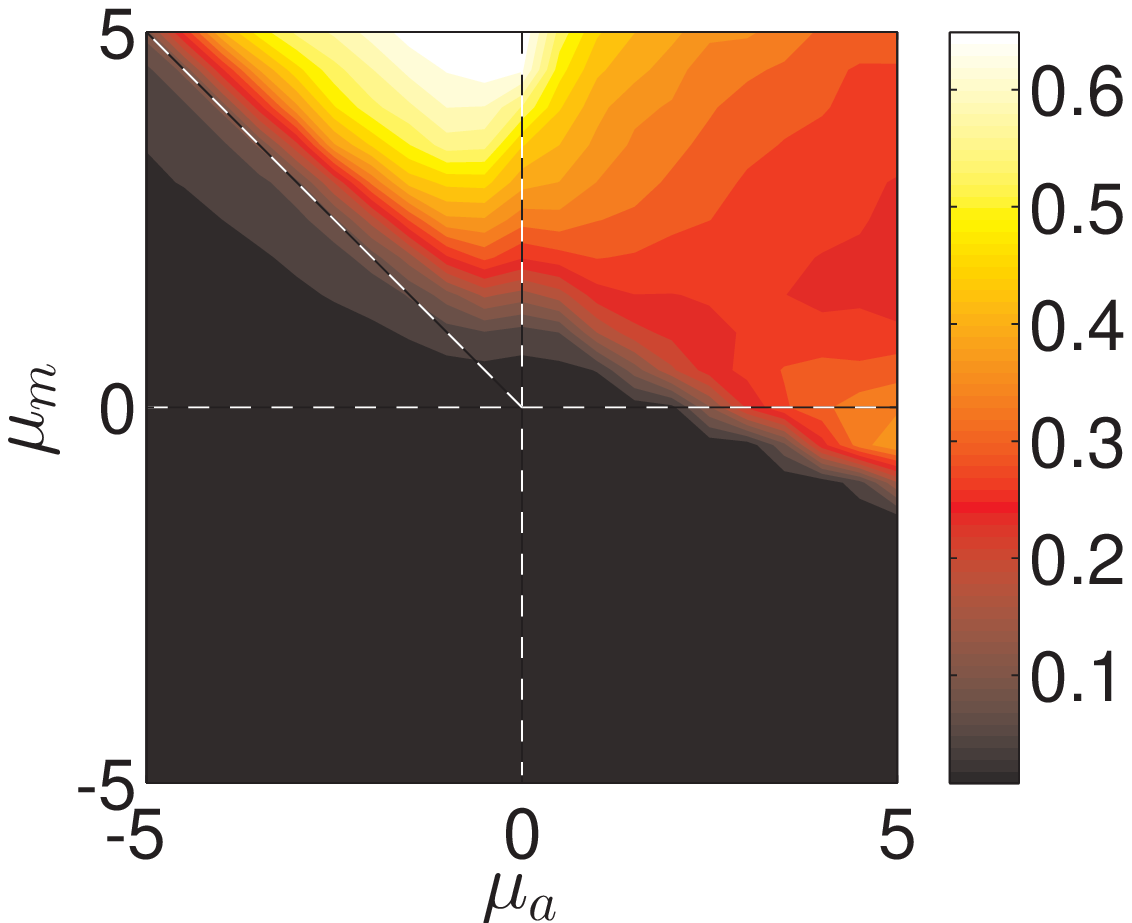}
\end{minipage}
{\large $\rho_s=1.25$, $D_\varphi=0.1$ \vspace{1ex} \\} 
\begin{minipage}{0.47\linewidth}
\centering center of mass speed $\langle S \rangle_t$ \\
\includegraphics[width=0.8\linewidth]{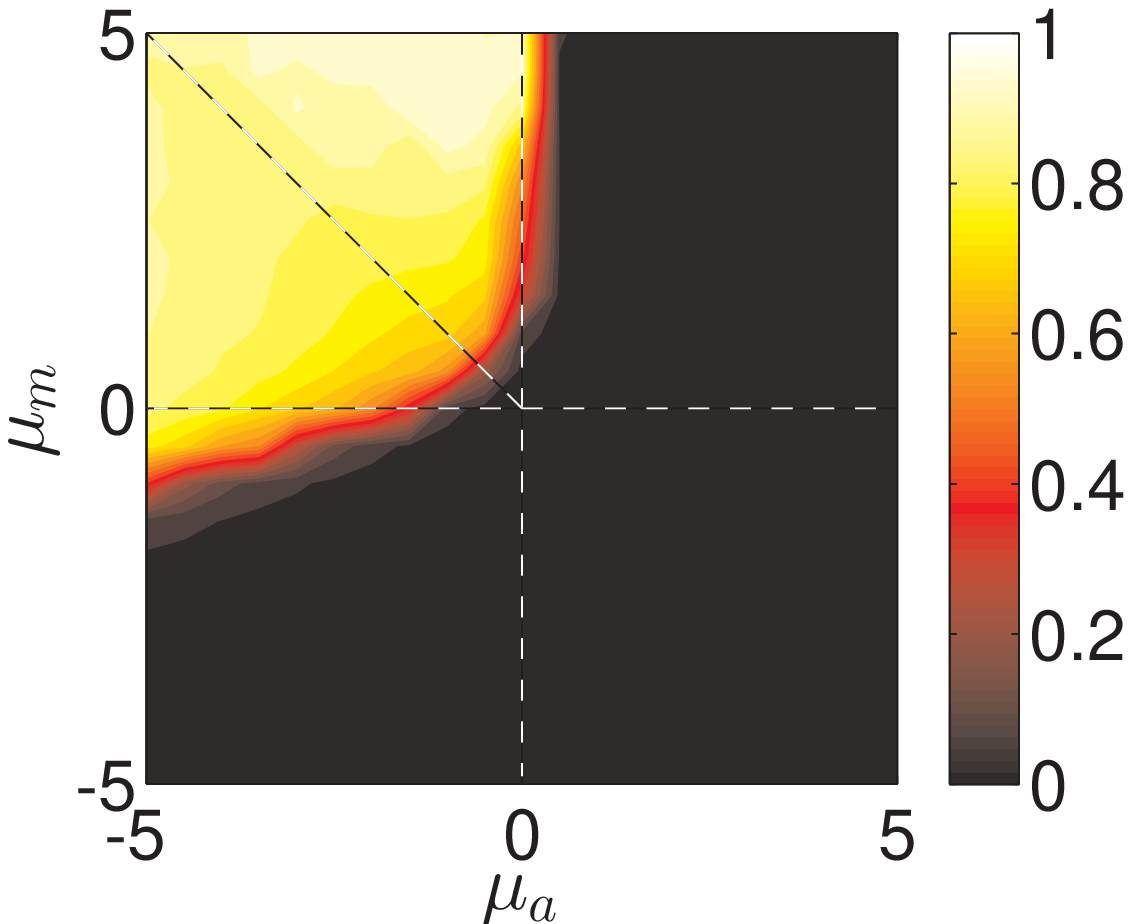}
\end{minipage}
\begin{minipage}{0.47\linewidth}
\centering  neighbor number $\langle N \rangle_t$ \\
\includegraphics[width=0.8\linewidth]{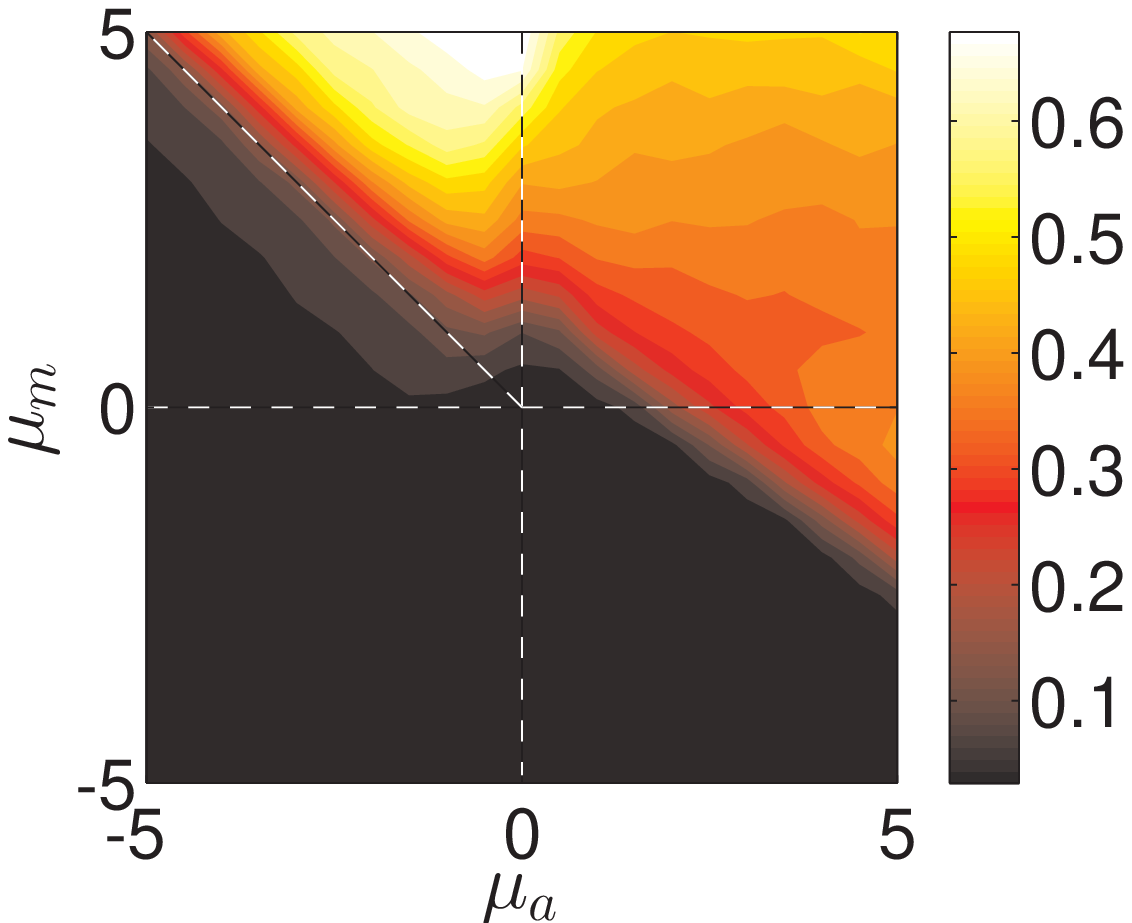}
\end{minipage}
{\large $\rho_s=5.00$, $D_\varphi=0.1$ \vspace{1ex} \\} 
\begin{minipage}{0.47\linewidth}
\centering center of mass speed $\langle S \rangle_t$ \\
\includegraphics[width=0.8\linewidth]{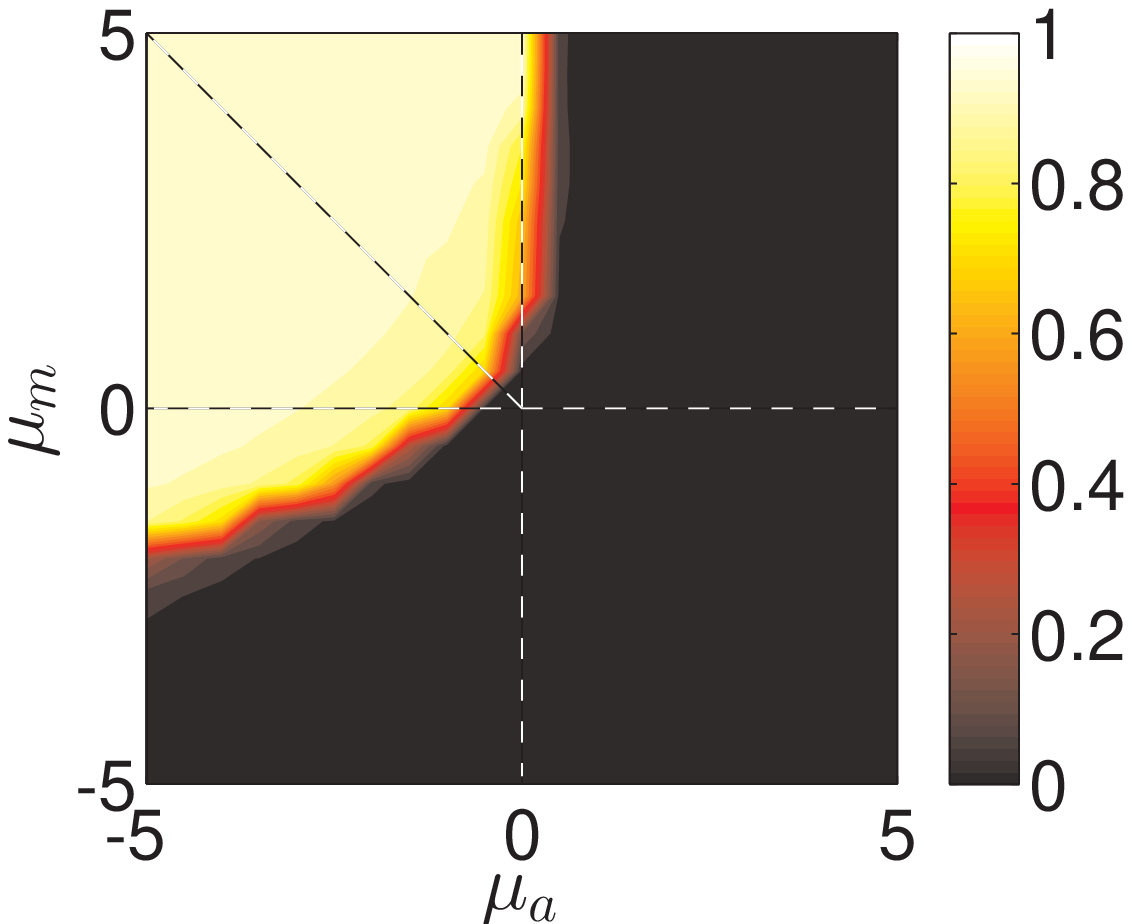}
\end{minipage}
\begin{minipage}{0.47\linewidth}
\centering  neighbor number $\langle N \rangle_t$ \\
\includegraphics[width=0.8\linewidth]{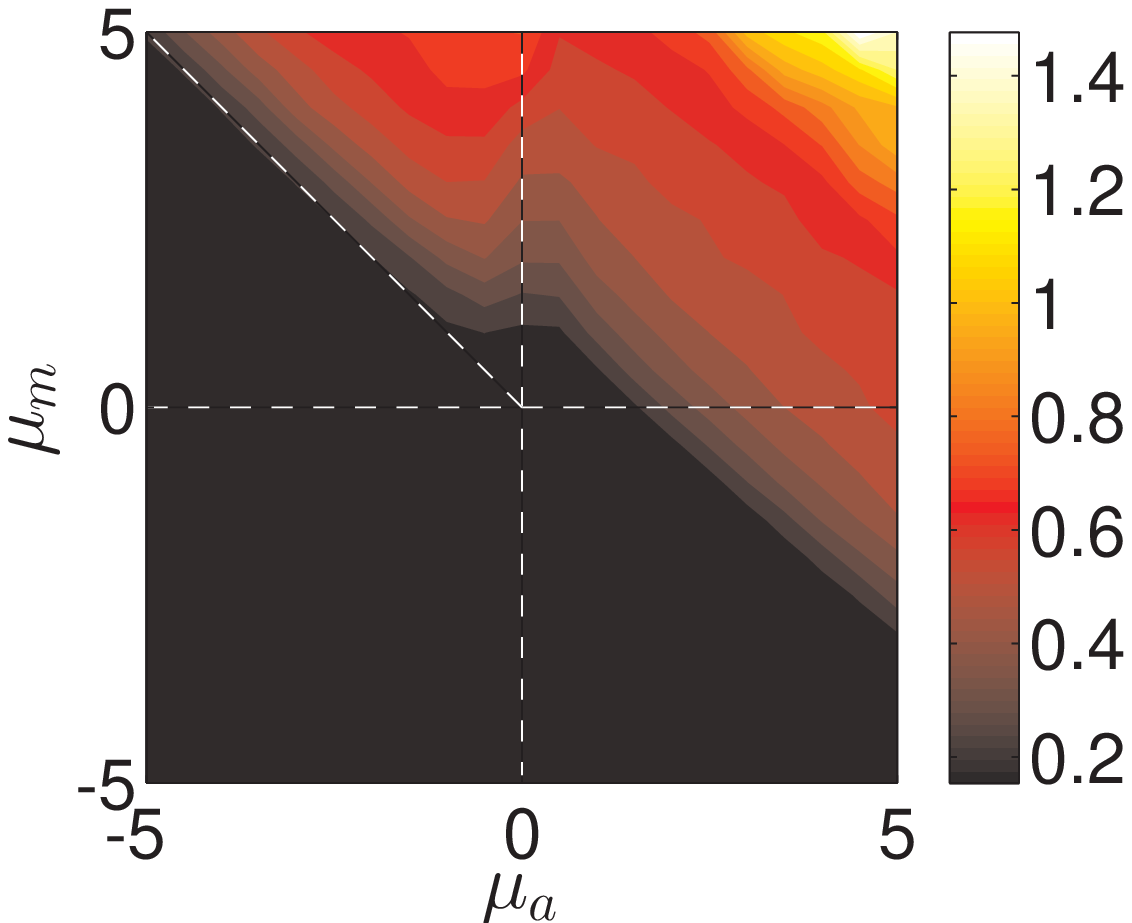}
\end{minipage}
\caption{Steady-state center-of-mass speed $\langle S \rangle_t$ (left) and the neighbor number $\langle N \rangle_t$ (right) versus $\mu_m$ and $\mu_a$ for different densities $\rho_s=0.56$ (top), $1.25$, $5.00$ and $D_\varphi=0.1$. The vertical and horizontal dashed lines indicate the zero axes. The diagonal dashed-line in the escape and pursuit quadrant indicates the border between escape-dominated (below the diagonal) and pursuit-dominated behavior (above the diagonal). \label{fig:const_speed_diffR_Dp01}}
\end{center}
\end{figure}

\begin{figure}
\begin{center}
{\large $\rho_s=1.25$, $D_\varphi=0.02$  \vspace{1ex} \\} 
\begin{minipage}{0.47\linewidth}
\centering center of mass speed $\langle S \rangle_t$ \\
\includegraphics[width=0.8\linewidth]{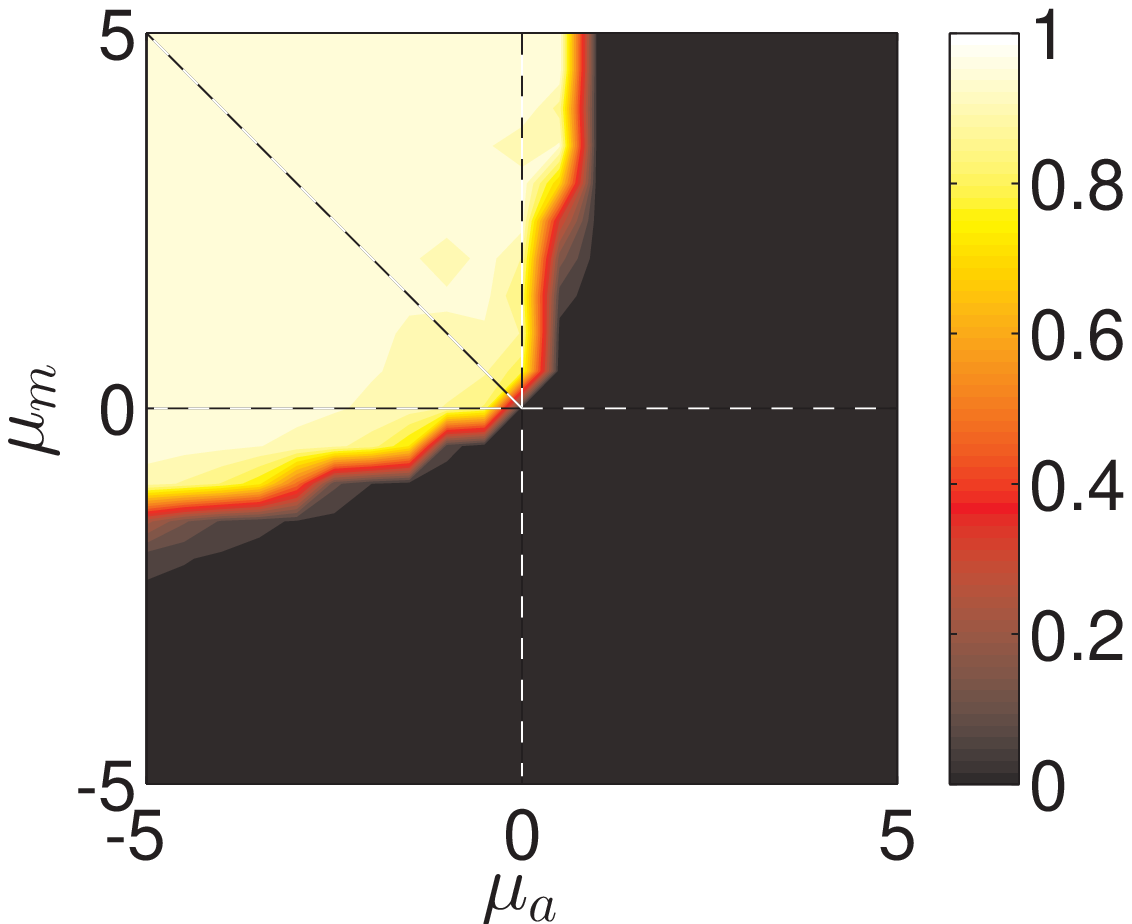}
\end{minipage}
\begin{minipage}{0.47\linewidth}
\centering  neighbor number $\langle N \rangle_t$ \\
\includegraphics[width=0.8\linewidth]{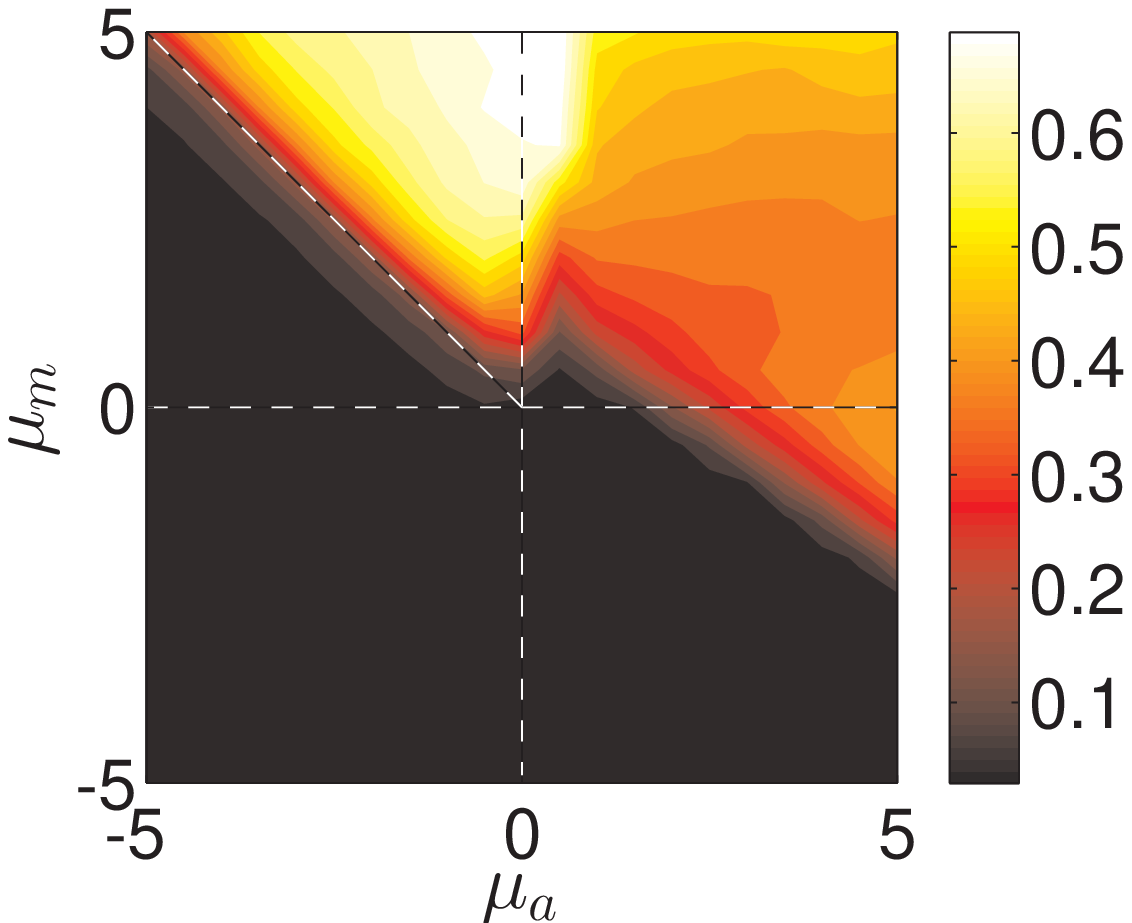}
\end{minipage}
{\large $\rho_s=1.25$, $D_\varphi=0.5$ \vspace{1ex} \\} 
\begin{minipage}{0.47\linewidth}
\centering center of mass speed $\langle S \rangle_t$ \\
\includegraphics[width=0.8\linewidth]{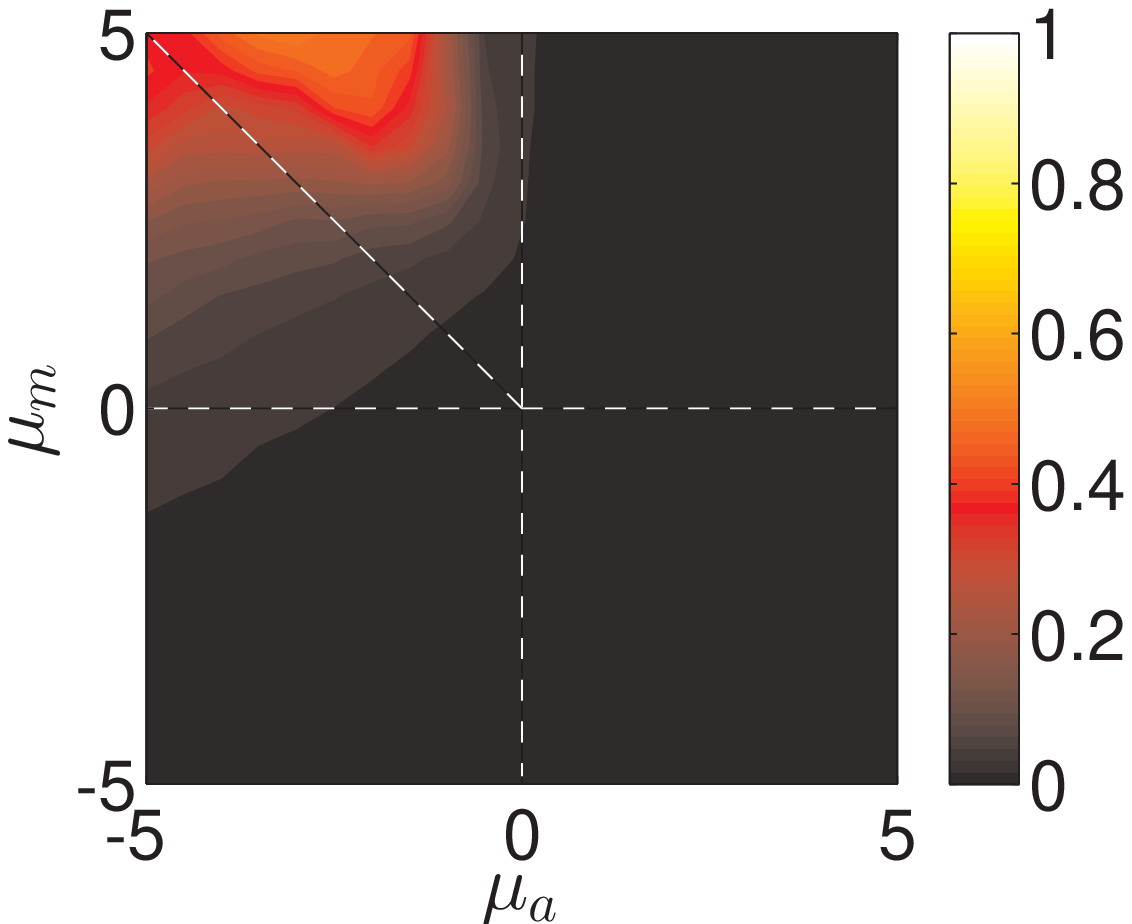}
\end{minipage}
\begin{minipage}{0.47\linewidth}
\centering  neighbor number $\langle N \rangle_t$ \\
\includegraphics[width=0.8\linewidth]{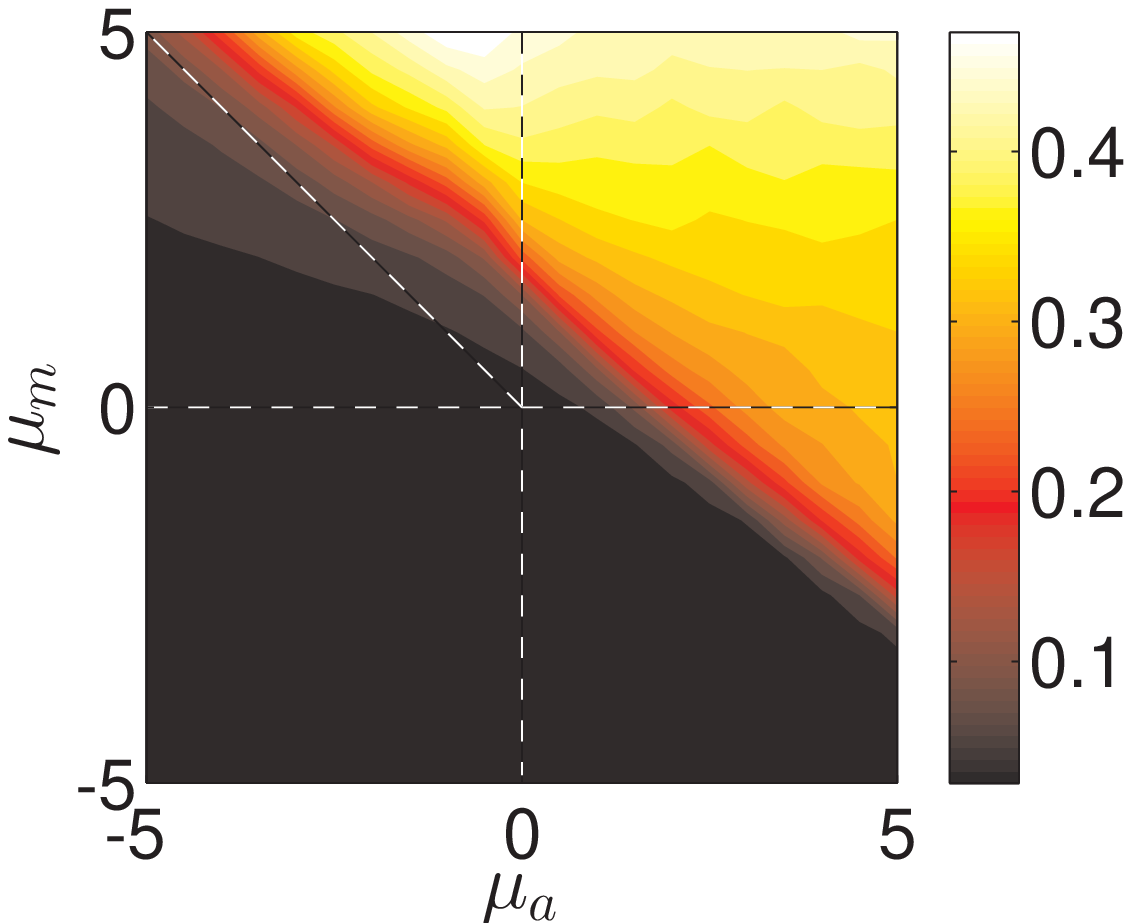}
\end{minipage}
{\caption{Steady-state center-of-mass speed $\langle S \rangle_t$ (left) and the neighbor number $\langle N \rangle_t$ (right) versus $\mu_m$ and $\mu_a$ for low angular noise $D_\varphi=0.02$ (top) and high  angular noise $D_\varphi=0.5$ (bottom). The dashed lines as in Fig. \ref{fig:const_speed_diffR_Dp01}. \label{fig:const_speed_R1250_diffDp}}}
\end{center}
\end{figure}

Throughout this work we set $s_0=1$, $l_r=1$ and $l_s=5$. Furthermore, we use $\mu_r=20$, which ensures that for binary interactions the short-ranged repulsion is always larger than the sum of the other possibly attracting forces. The particle number is set constant to $N=2000$ and the density is varied by changing the system size $L$. The parameter-space diagrams in Figs. \ref{fig:const_speed_diffR_Dp01} and \ref{fig:const_speed_R1250_diffDp} were obtained from interpolating the results for $\left\langle S \right\rangle_t$ and $\langle N \rangle_t$ for  $441$ ($21\times21$) individual, evenly spaced, grid points in the interaction parameter space with $-5.0\leq \mu_m,\mu_a \leq +5.0$. Each such point corresponds to an average over the results of 6 independent simulation runs, whereby for each run a temporal average was taken after the system reached a steady-state.

At sufficiently high densities and sufficiently low (angular) noise, we can observe the onset of collective motion for a wide range of interaction parameters (see Figs. \ref{fig:const_speed_diffR_Dp01} and \ref{fig:const_speed_R1250_diffDp}). At low noise and high density, the region of collective motion coincides approximately with the escape and pursuit quadrant of the interaction parameter space with $\mu_m>0$ and $\mu_a<0$. It contains the special cases of only attraction to moving away (repulsion from approaching) individuals with $\mu_m>0$ and $\mu_a=0$ ($\mu_a<0$ and $\mu_m>0$), and extends also into the pure repulsion region ($\mu_a,\mu_m<0$ and $\mu_a\ll\mu_m$) and to a much lesser extent into the pure attraction region ($\mu_a,\mu_m>0$ and $\mu_m\ll\mu_a$). This is agreement with the predictions of the kinetic theory for the monopole approximation the effective social interaction. However, the region of collective motion is smaller in simulations than in the simple theory. This may be due to the impact of density inhomogeneities, finite short-range repulsion, and/or higher order effects.  

Within the escape and pursuit regime, where the repulsion to approaching individuals (escape) dominates over the attraction to moving away individuals (pursuit), the neighbor number is low. This corresponds to low degree of clustering and a rather homogeneous spatial distribution of particles throughout the system. The neighbor number $\langle n \rangle$ increases strongly in the pursuit dominated regime ($|\mu_m|>|\mu_a|$ with $\mu_m>0$, $\mu_a<0$) indicating strong density inhomogeneities corresponding to dense collectively moving bands and clusters. This resembles the behavior observed in the original Brownian particle escape \& pursuit model.

\begin{figure}
\begin{center}
\centering\includegraphics[width=\linewidth]{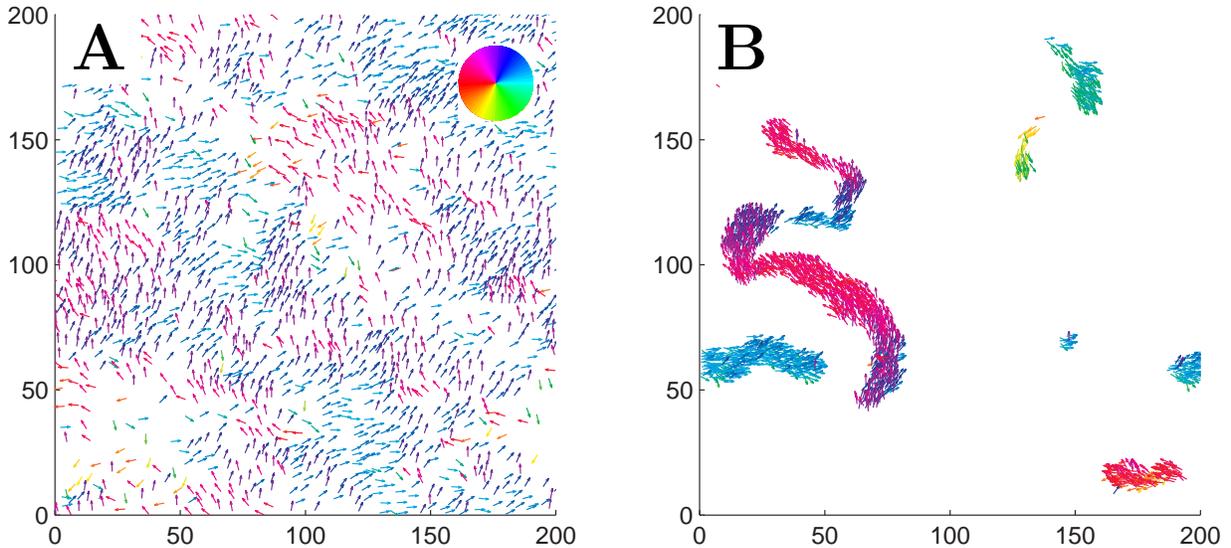}
\caption{Examples of spatial configurations for the pure escape case (A) with $\mu_a=-3$ and $\mu_m=0$ and the pure pursuit case (B) with$\mu_a=0$ and $\mu_m=+3$. \label{fig:epspp_snapshots_eo_po}
}
\end{center}
\end{figure}

The neighbor number is also high in the pure attraction regime ($\mu_a,\mu_m>0$) without collective motion, where clusters with vanishing center of mass velocity can be observed. 
Interestingly, at moderate densities ($\rho_s=0.56$, $1.25$ in Figs. \ref{fig:const_speed_diffR_Dp01},\ref{fig:const_speed_R1250_diffDp}), 
the maximum of the neighbor number is located in the pursuit-dominated regime with collective motion, and not, as one might expect, in the regime of (strong) overall attraction ($\mu_m,\mu_a\gg0$). 
In the ordered state, particles move approximately in the same direction and the relative speed $|\tilde v_{ji}|$ will be close to zero. For low repulsion from approaching individuals the escape response is negligible, whereas the attraction to moving away individual suffices to maintain cohesion, in particular at low noise strengths.   Effectively, the density of such collectively moving cluster is limited by the short-range repulsion (see Fig. \ref{fig:epspp_snapshots_eo_po}B). In the pure attraction regime, particles on the boundaries of a cluster will be attracted towards the local center of mass. However, due to the self-propelled motion with inertia and scattering with other individuals within the disordered cluster they will eventually move outwards again. As a result we observe disordered aggregates, which resemble mosquito swarms (see Fig.  \ref{fig:epspp_snapshots}B), and are more dilute in comparison to the coherently moving clusters in the pursuit-dominated case. 

As might be expected, increased stochasticity in the motion of individuals, inhibits the onset of collective motion. The region of parameter space with $\langle S \rangle_t$  significantly larger than $0$ reduces strongly with increasing $D_\varphi$ by receding towards the regime of strong escape and pursuit response (see Fig \ref{fig:const_speed_R1250_diffDp}) in agreement with the kinetic theory.      

For the Head-on-Head regime as well as for pure repulsion (with $\mu_m\gg0$) a quasi homogeneous distribution of particles can be observed with no collective motion (see Figs. \ref{fig:epspp_snapshots}C,D and \ref{fig:const_speed_diffR_Dp01},\ref{fig:const_speed_R1250_diffDp} ).

\section{Discussion}

In this work, we have analyzed  a model for collective dynamics based on selective attraction and repulsion interactions, which was recently used to model the evolution of phenotypic phase change in locusts \cite{guttal_cannibalism_2012}. The modelling of individual dynamics in terms stochastic differential equation (Langevin equations), allows a straight forward derivation of a kinetic description, which may be used for further theoretical analysis based on mean-field considerations and moment expansion of the corresponding probability density function \cite{romanczuk_mean-field_2011,grossmann_active_2012}.  

The model is able to account for three types of social responses relevant in the biological context: escape and pursuit, pure avoidance and pure attraction behavior. We have shown that large-scale collective motion of self-propelled particles emerge without any explicit velocity-alignment mechanism in the (generalized) escape and pursuit regime, without the spatial anisotropy in the social interaction in the original escape and pursuit model \cite{romanczuk_collective_2009}. Here, we should note that in the original escape-pursuit model, with Brownian dynamics of individual agents, the spatial anisotropy is essential for the emergence of directed collective motion.

In general, the spatial distribution of individuals during collective motion depends strongly on the relative strength of the different social forces. If escape dominates we observe a homogeneous spatial distribution (Fig. \ref{fig:epspp_snapshots_eo_po}A), whereas in the pursuit-dominated case compact, coherently moving structures, as for example snake-like clusters, can be observed (Fig. \ref{fig:epspp_snapshots_eo_po}B). In between, for comparable escape and pursuit strengths the band like structures perpendicular to the average direction of motion emerge, which appear also in systems with velocity-alignment (Fig. \ref{fig:epspp_snapshots}A) \cite{gregoire_onset_2004,romanczuk_active_2012}. 

The region of collective motion decreases with increasing noise as well as with decreasing density. However, at low densities, the region of collective motion shows a clear shift towards the pursuit-dominated regime (see Fig \ref{fig:const_speed_diffR_Dp01} top), where in a finite system a order parameter is maintained by relatively few moving clusters containing most of the individuals.  

These results are not only in agreement with the basic, qualitative predictions, drawn from the kinetic theory in Sect. \ref{sec:kinetic}, but resemble also the qualitative behavior of the original escape-pursuit model \cite{romanczuk_collective_2009}. Interestingly, at intermediate densities and within the interaction range studied, the maximum of the neighbor number -- indicating strongest clustering in the system -- appears in the pursuit-dominated regime and not for pure attraction. This can be understood from the fact that local order decreases the effective ``temperature'' associated with absolute deviations of the velocities of single particles from the average velocity of their neighbors\cite{romanczuk_collective_2010,romanczuk_mean-field_2011}. As a result we observe a decrease in the active pressure due to the stochastic self-propelled nature of individual motion, which counteracts the concentration of individuals due to attractive forces.

In conclusion, the modelling of collective motion in biology via selective attraction-repulsion interactions appears very promising. The model accounts for various individual behaviors 
and displays different spatial patterns of collective motion. 
The response based only on the distinction between approaching and moving away individuals can be directly linked to the response to looming visual stimuli, which has been shown to play an important role in various species \cite{schiff_persistent_1962,rind_orthopteran_1992,rind_orthopteran_1992,rind_orthopteran_1992,devries_loom-sensitive_2012}. Finally, in this context, we should mention a recent work by Lemasson and coworkers \cite{lemasson_collective_2009}. They introduce a model for collective motion based on selective interaction of individuals based on explicit, simplified description of visual information available to each individual. 

\section*{Acknowledgment}

PR would like to thank Vishwesha Guttal (Indian Institute of Science) and Iain D. Couzin (Princeton University) for many helpful discussions on the subject. 
Part of the work was performed during a stay of LSG at IFISC in Palma de Mallorca. LSG thanks for the great hospitality and the fruitful cooperation. Furthermore, LSG
acknowledges the support by the DFG via the IRTG 1740.

%\clearpage
%\bibliographystyle{unsrt}	
\bibliography{ColMot_Interface}

\end{document}